\documentclass[11pt,a4paper,english]{article}
\pdfoutput=1
\usepackage[english]{babel}
\usepackage[utf8]{inputenc}
\usepackage[left=0.8in,right=0.8in,top=1.0in,bottom=1.2in]{geometry}

\usepackage{amsmath,amssymb,amsfonts, bm,bbm,slashed}
\usepackage{graphicx}
\usepackage[sort&compress]{natbib}
\usepackage[normalem]{ulem}
\usepackage{enumerate}
\usepackage{epsfig, subfigure}
\usepackage{setspace}
\usepackage{booktabs, tabularx}
\usepackage{units}
\usepackage[utf8]{inputenc}
\usepackage{lineno}
\usepackage{url}

\usepackage{cancel}
\usepackage{hyperref}
\usepackage{cleveref}

\hypersetup{
	pdfnewwindow=true,   
	colorlinks=true,     
	linkcolor=blue,      
	citecolor=blue,      
	filecolor=blue,      
	urlcolor=blue        
}

\makeatletter
\g@addto@macro\bfseries{\boldmath}
\makeatother

\allowdisplaybreaks

\bibpunct{[}{]}{,}{n}{}{,}

\newcommand{\be} {\begin{equation}}
\newcommand{\ee} {\end{equation}}
\newcommand{\ba} {\begin{array}}
\newcommand{\ea} {\end{array}}

\newcommand{\gsim}{\lower.7ex\hbox{$\;\stackrel{\textstyle>}{\sim}\;$}}
\newcommand{\lsim}{\lower.7ex\hbox{$\;\stackrel{\textstyle<}{\sim}\;$}}

\newcommand{\mttot}{\ensuremath{m_\text{T}^\text{tot}}}
\newcommand{\GeV}{{\,\rm GeV}}

{\renewcommand{\arraystretch}{1.2}

\begin{document}

\begin{flushright}
 ZU-TH-19/05  \\
\end{flushright}

\begin{center}
\vspace{1.5cm}
     {\Large\bf High-$p_T$ Signatures in Vector--Leptoquark Models}
       \\ [1cm] 

   {\bf Michael J.~Baker,  Javier Fuentes-Mart\'{\i}n, Gino Isidori, Matthias K\"{o}nig}    \\[0.5cm]
  {\em Physik-Institut, Universit\"at Z\"urich, CH-8057 Z\"urich, Switzerland}  \\
\end{center}
\vspace{1cm}

\centerline{\large\bf Abstract}
\begin{quote}
We present a detailed analysis of the collider signatures of TeV-scale 
massive vector bosons motivated by the hints of lepton flavour non-universality 
observed in $B$-meson decays. We analyse three 
representations that necessarily appear together in a large class of 
ultraviolet-complete models:  a  colour-singlet ($Z'$),  
a colour-triplet (the $U_1$ leptoquark), and a colour octet ($G'$).
Under general assumptions for the interactions of these exotic states with Standard Model fields,
including in particular possible right-handed and flavour off-diagonal couplings for the $U_1$,
we derive a series of stringent bounds on masses and couplings that constrain 
a wide range of explicit new-physics models.
\end{quote}

\section{Introduction}
\label{sec:intro}

The hints of Lepton Flavour Universality (LFU) violation in semi-leptonic $B$ decays,
namely the deviations from  $\tau/\mu$ (and $\tau/e$) universality in
 $b\to c \ell \bar\nu$ decays~\cite{Lees:2013uzd,Aaij:2015yra,Hirose:2016wfn,Aaij:2017deq}
 and the deviations from $\mu/e$ universality in  $b\to s \ell \bar{\ell}$ decays~\cite{Aaij:2014ora,Aaij:2017vbb},
are among the most interesting departures from the Standard Model (SM)
reported by experiments in the last few years.
The attempt to find a single beyond-the-SM (BSM) explanation for the combined set of anomalies 
has triggered intense theoretical activity, whose interest goes beyond
the initial phenomenological motivation. In fact, it has shed light on new classes of SM extensions 
that turn out to be very interesting {\em per se}  
and that have not been investigated in great detail so far,
pointing to non-trivial dynamics at the TeV scale
possibly linked to a solution of the SM flavour puzzle.

The initial efforts to address both sets of anomalies have been focused on Effective Field Theory (EFT) 
approaches via four-fermion effective operators 
(see~\cite{Bhattacharya:2014wla, Alonso:2015sja, Greljo:2015mma, Calibbi:2015kma} for the early attempts).
However, the importance of complementing EFT 
approaches with appropriate simplified models with new 
heavy mediators was soon realised~\cite{Greljo:2015mma, Barbieri:2015yvd}.
Given the relatively low scale of new physics hinted by the charged-current anomalies, the impact of
considering a full model rather than an EFT on high-$p_T$ 
constraints are significant~\cite{Faroughy:2016osc,Greljo:2017vvb,Greljo:2018tzh}.
More recently, a further advancement has been achieved with the development of 
more complete (and more complex) models with a consistent ultraviolet (UV)  behaviour
(see in particular~\cite{DiLuzio:2017vat, Assad:2017iib, Calibbi:2017qbu,  Bordone:2017bld, Barbieri:2017tuq, Greljo:2018tuh, Blanke:2018sro, DiLuzio:2018zxy, Marzocca:2018wcf, Kumar:2018kmr, Becirevic:2018afm,Bordone:2018nbg,Fornal:2018dqn}).

In early EFT attempts, it was realised that a particularly good mediator accounting 
for both sets of anomalies is a TeV-scale $U_1\sim(\mathbf{3},\mathbf{1},2/3)$ vector leptoquark, coupled mainly 
to third-generation fermions~\cite{Alonso:2015sja,Barbieri:2015yvd}. The effectiveness of this state as 
a single mediator accounting for all available low-energy data has been clearly established in~\cite{Buttazzo:2017ixm}. 
However, this state can not be the only TeV-scale vector particle in a realistic extension of the SM.
Since it is a massive vector, the $U_1$ can be either a massive gauge boson of a spontaneously broken 
gauge symmetry $G_{\rm NP} \supset G_{\rm SM}$, as in the attempts proposed in~\cite{DiLuzio:2017vat, Assad:2017iib, Calibbi:2017qbu, Bordone:2017bld},
or a vector resonance of some new strongly interacting dynamics, as e.g.~in~\cite{Barbieri:2017tuq, Blanke:2018sro}.
As we show, in both cases the consistency of the theory requires additional vector states with similar masses.
The purpose of this paper is to provide a comprehensive analysis of the high-$p_T$ constraints on 
the vector leptoquark $U_1$ and what can be considered its minimal set of vector companions, namely a colour octet
$G^\prime\sim(\mathbf{8},\mathbf{1},0)$, which we will refer to as the coloron, and a colour singlet $Z^\prime\sim(\mathbf{1},\mathbf{1},0)$. 

In our analysis we consider the most general chiral structure 
for the $U_1$ couplings to SM fermions. This is in contrast with many recent studies which considered only 
 left-handed (LH) couplings. While this hypothesis is motivated by the absence of clear indications 
of right-handed (RH) currents in the present data and by the sake of minimality, it does not have a strong theoretical justification. 
Indeed, the quantum numbers of the $U_1$ allow for RH couplings, 
and in motivated UV completions such couplings naturally appear~\cite{Bordone:2017bld,Bordone:2018nbg}.
We also analyse the impact of a non-vanishing mixing between the second and third family in high-$p_T$ searches,
including in particular constraints from $pp\to \tau\mu$ and $pp\to \tau\nu$.
As we show, the inclusion of right-handed couplings and/or a sizeable 2-3 family mixing yields 
significant modifications to the results found in the existing literature.

The structure of this paper is as follows: In~\cref{sec:apero} we motivate our choice of TeV-scale vectors 
and in \cref{sec:model} we introduce the phenomenological Lagrangian adopted to describe their  high-$p_T$
signatures. We then present the results of the searches in \cref{sec:results} and conclude with \cref{sec:conclusions}.

	\section{The spectrum of vector states at the TeV scale}
	\label{sec:apero}	

The bottom-up requirement for the class of models we are interested in 
is the following effective interaction of the $U_1$ field with SM fermions:
\begin{align}
&\mathcal{L}_{\rm NP} \supset  \frac{g_U}{\sqrt{2}}\, U_1^{\mu,\alpha}  \left[ (J^L_U)_{\mu}^{\alpha} +   (J^R_U)_{\mu}^{\alpha}\right] +{\rm h.c.}~, 
\nonumber \\ 
 &(J^L_U)_{\mu}^{\alpha} =  \beta_L^{ij}\,\bar q^{i,\alpha}_L\gamma_\mu\ell_L^j~, \quad 
       (J^R_U)_{\mu}^{\alpha}=  \beta_R^{ij}\,\bar d^{i,\alpha}_R\gamma_\mu e_R^j~.
 \label{eq:U1curr}
\end{align}
Here $q_L\, (\ell_L)$ denotes the left-handed quark (lepton) doublets, $d_R\, (e_R)$ denotes the right-handed down-type quark (charge-lepton) singlets,
$i \in \{1,2,3\}$ and $j \in \{1,2,3\}$ are flavour indices, $\alpha \in\{ 1,2, 3\}$ is a $SU(3)_c$ index, and  $\beta_{L,R}^{ij}$ are complex matrices in family-space.

The effective interaction in \cref{eq:U1curr} unambiguously identifies the representation of 
$U_1$ under $G_{\rm SM} = SU(3)_c \times SU(2)_L \times U(1)_Y$ to be  $(\mathbf{3},\mathbf{1},2/3)$.
There are two basic classes of well-defined UV theories where such interactions can occur:
\begin{enumerate}
\item[i.] {\em Gauge models}. Here $U_1$ is the massive gauge boson of a spontaneously broken 
gauge symmetry $G_{\rm NP} \supset G_{\rm SM}$.  The need for extra massive vectors follows from 
the size of the coset-space of  $G_{\rm NP}/G_{\rm SM}$, that necessarily requires additional generators
besides the six associated to $U_1$.
\item[ii.]
 {\em Strongly interacting models}. Here 
$U_1$ appears as a massive resonance for a new strongly interacting sector. In this case the need of 
additional massive vectors is a consequence of the additional resonances formed by the same set of 
constituents leading to $U_1$. 
\end{enumerate}

\subsubsection*{Gauge models: the need for a \texorpdfstring{$Z^\prime$}{Z'}}

Within gauge models, let us start analysing the case of a single generation of SM fermions ($i=j=3$), 
and further assume that SM fermions belong to well-defined representations of $G_{\rm NP}$ 
(i.e.~no mixing between SM-like and exotic fermions). Under these assumptions, $\beta_L$ is non-zero
only if $q_L$  and $\ell_L$ belong to the same $G_{\rm NP}$ representation. We denote this 
representation $\psi_L$ and, without loss of generality, we decompose it as
\begin{align}
\psi_L =  \psi_L^{\rm SM} + \psi_L^{\rm exotic}~, \qquad  \psi^{\rm SM}_L = \left( \ba{c} q^\beta_L \\ \ell_L \ea \right)~.
\label{eq:psiBL}
\end{align}
In this notation the left-handed current in \cref{eq:U1curr} can be written as 
$(J^L_U)_{\mu}^\alpha =   \bar\psi^{\rm SM}_L   (T^\alpha_+)  \gamma_\mu  \psi^{\rm SM}_L$
with the following explicit expression for the action of the $G_{\rm NP}$
generators on the SM projection of $\psi_L$:
\begin{equation}
T^\alpha_+  = \left( \ba{cc} 0 & \delta_{\alpha\beta} \\  0 &  0 \ea\right)~.   
\end{equation}
The closure of the algebra of the six generators $T^\alpha_\pm$ associated with the six components of $U_1$
implies the need of the following additional (colour-neutral) generator
\begin{equation}
T_{B-L} = \left( \ba{cc} \frac{1}{3}\delta_{\beta\gamma} & 0  \\  0 &  -1 \ea \right)~, \qquad 
\frac{1}{3} \sum_{\alpha,\delta=1}^3  [  T^\alpha_+, T^\delta_- ] =  T_{B-L}~. 
\end{equation}
The same conclusion is reached by looking at the right-handed coupling in \cref{eq:U1curr}.
Moreover, since a possible mixing between SM and exotic fermions must occur
in a $SU(3)_c$ invariant way, the decomposition in \cref{eq:psiBL} also holds for possible 
exotic fermions mixing with the SM ones.
Hence the need of $T_{B-L}$ for the closure of the 
algebra is a general conclusion that holds independently of the possible mixing 
among fermion representations. 

An equivalent way to deduce the need for an extra generator  is the
observation that the minimal group $G^{\rm min}_{\rm NP} \supset G_{\rm SM}$
containing generators associated to the representation $(\mathbf{3},\mathbf{1},2/3)$ is 
\begin{equation}
G_{\rm NP}^{\rm min} = SU(4) \times SU(2)_L \times U(1)_{T^3_R}~,
\label{eq:Gmin}
\end{equation}
i.e.~the subgroup of the Pati-Salam group 
$G_{\rm PS} = SU(4) \times SU(2)_L \times SU(2)_R$~\cite{Pati:1974yy}. 
$G_{\rm NP}^{\rm min}$ is obtained by considering the $U(1)$ subgroup of 
$SU(2)_R$ defined by its diagonal (electric-charge neutral) generator $T^3_R$.
The coset $G^{\rm min}_{\rm NP}/G_{\rm SM}$ contains seven generators: the six $T^\alpha_\pm$
describing the coset $SU(4)/SU(3)_c\times U(1)_{B-L}$,  and $T_{B-L}$.

In gauge models, the presence of an extra massive vector 
$Z^\prime\sim(\mathbf{1},\mathbf{1},0)$ associated with the breaking 
$U(1)_{B-L} \times U(1)_{T^3_R} \to U(1)_Y$ is thus unavoidable. Since 
the breaking of $U(1)_{B-L}$ necessarily implies a breaking of $SU(4)$,  
the breaking terms which lead to a non-vanishing $Z^\prime$ mass 
necessarily induce a mass term for the $U_1$ as well. Hence, the $Z^\prime$ state cannot be decoupled. 
The opposite is not true: since the $U_1$ generators are associated to the 
$SU(4)/SU(3)_c\times U(1)_{B-L}$ coset, mass terms for the $U_1$ do not necessarily contribute 
to the $Z^\prime$ mass.

\subsubsection*{Gauge models: the need for a \texorpdfstring{$G^\prime$}{coloron}}
While the minimal group in \cref{eq:Gmin} allows us to build a consistent model for a massive 
$U_1\sim (\mathbf{3},\mathbf{1},2/3)$, it does not leave us enough freedom to adjust 
$U_1$ and $Z^\prime$ couplings in order to comply with low- and high-energy data.

Under $G_{\rm NP}^{\rm min}$ the interaction strengths of both $U_1$ and $Z^\prime$
are unambiguously related to the QCD coupling ($g_s$) and to hypercharge, 
given that they all originate from the same $SU(4)$ group.  In particular $g_U =g_s (M_{U_1})$,  
in a normalisation where $|\beta_{L,R}^{ij}| \leq 1$. Moreover, the couplings of the $Z^\prime$ 
to SM fermions are necessarily flavour universal.\footnote{This statement  follows from the fact that 
the mixing of SM fermions among themselves (in flavour space) and with possible exotic representations
necessarily involve states with the same $B-L$ charges. As a result, the mixing acts as a unitary rotation 
on the $Z^\prime$ couplings that remains proportional to the identity matrix in flavour space.}
A flavour-universal $Z^\prime$  
is constrained by LHC dilepton searches to have $M_{Z^\prime} \gsim 5$~TeV~\cite{Sirunyan:2018exx,  Aaboud:2017buh}.
Within $G_{\rm NP}^{\rm min}$, 
the $U_1$ should be necessarily close in mass~\cite{DiLuzio:2018zxy} which,
 together with the low value of $g_U$, results in a negligible impact on  $b\to c \ell\nu$ decays.

To avoid these constraints, $T^\alpha_\pm$, $T_{B-L}$, and the QCD generators $T^a$, should not be 
unified in a single $SU(4)$ group. Given the commutation rules between $T^\alpha_\pm$ and $T^a$,
the next-to-minimal option is obtained with~\cite{DiLuzio:2017vat}
\begin{align}
(G_{\rm NP}^{\rm min})^\prime = SU(4) \times  SU(3)^\prime \times SU(2)_L \times U(1)_{T^3_R}\,,
\label{eq:Gntomin} 
\end{align}
where  $SU(3)_c$ is the diagonal subgroup of $SU(4) \times  SU(3)^\prime$
(see also \cite{Georgi:2016xhm,Diaz:2017lit}). In this case we can achieve the two goals of 1) 
decoupling the overall coupling 
of $U_1$ from $g_c$, letting it reach the higher values needed to impact 
$B$-physics data with $M_{U_1}\sim$ few TeV; 2) having flavour non-universal couplings 
for both $U_1$ and $Z^\prime$. The latter can be achieved either via mixing 
with exotic fermions (as in~\cite{DiLuzio:2017vat}), and/or with a flavour-dependent
assignment of the $SU(4)\times SU(3)^\prime$ quantum numbers (as in~\cite{Bordone:2017bld,Greljo:2018tuh}).

The enlargement of the coset space to $(G_{\rm NP}^{\rm min})^\prime/G_{\rm SM}$ 
directly requires a massive colour-octet vector (the ``coloron'' $G^\prime$) associated to the 
breaking $SU(3)_{[4]} \times  SU(3)^\prime$, where  
$SU(3)_{[4]}$ is the ``coloured" subgroup of  $SU(4)$. Similarly to the case of the $Z'$, 
breaking terms leading to a non-vanishing $G^\prime$ mass necessarily induces 
a mass term also for the $U_1$, while the opposite is not necessarily true. 

\subsubsection*{Vector spectrum in strongly interacting models}
In strongly interacting models, the leptoquark $U_1$ is a composite state with two elementary fermions
charged under the new confining group $G_{\rm strong}$ as constituents. 
These fermions are necessarily charged under $SU(3)_c$ in order to 
generate a colour-triplet state.

The simplest option is the case of a vector triplet ($\chi_q^{\alpha}$) and a vector singlet ($\chi_\ell$),
both in the fundamental of $G_{\rm strong}$, such that
\begin{align}
\langle  0 |  \bar \chi_q^{\alpha} \gamma_\mu \chi_\ell | U_1 (p, \epsilon) \rangle = F_{U}  \epsilon_{\mu}^{\alpha}~,
\end{align}
where we have not explicitly indicated the $G_{\rm strong}$ indices. 
With these basic constituents one expects also one $G'$ and two $Z'$:
\begin{align}
\langle  0 |  \bar \chi_q^{\alpha} T^a_{\alpha\beta} \gamma_\mu \chi_q^\beta | G (p, \epsilon) \rangle = F_{G^\prime}  \epsilon_{\mu}^{a}~,
\quad  \langle  0 |  \bar \chi_q^{\alpha}  \gamma_\mu \chi_q^\alpha | Z_q^\prime (p, \epsilon) \rangle = F_{Z_q^\prime}  \epsilon_{\mu}~,
\quad \langle  0 | \bar \chi_\ell \gamma_\mu \chi_\ell | Z_\ell^{\prime} (p, \epsilon) \rangle = F_{Z_\ell^\prime}  \epsilon_{\mu}~.
\end{align}
The masses of these states are not precisely related to that of the $U_1$ as in the case of gauge models, 
but they are expected to be of similar size since they originate from the same dynamics. 
In principle one can enlarge the multiplicity of the constituents, e.g.~the colour triplet $U_1$ can be achieved by 
combining ${\bf 3}$ and $\bf {8}$ of $SU(3)_c$, but this can only increase the number 
of extra coloured vectors. A further exotic option is to consider $U_1$ as 
a fermion bilinear in an antisymmetric combination of $G_{\rm strong}$, as allowed e.g.~in $SU(2)$.
However, beside this peculiar case where symmetric $G_{\rm strong}$ combinations 
are forbidden (or much heavier in mass), this does not prevent the presence of at least one  $G'$ and one $Z'$
with masses comparable to the $U_1$.

	\section{Phenomenological Lagrangian}
	\label{sec:model}	

Having motivated the minimal set  $\{G',Z',U_1\}$ 
of massive vectors for a meaningful description of TeV scale dynamics, we 
proceed to set up a versatile framework for analysing the high-$p_T$ signatures 
of these states in a general way. In our analysis we restrict our attention to 
the interactions of these vectors with SM fermions and gauge bosons. We neglect 
possible Higgs couplings to the $Z^\prime$ since they are severely constrained 
by electroweak precision data (see e.g.~\cite{Buttazzo:2017ixm}) and are typically very small in the 
model realisations we are interested in. We also ignore any possible interactions 
of the extra vectors among themselves and to any other particles related to the 
UV completion of the model (either scalars or fermions). While some of the high-$p_T$ signatures
related to these interactions can be quite interesting~\cite{DiLuzio:2018zxy}, they are highly dependent on the details of the UV completion.  Here we only consider their possible indirect effects on the widths of the vectors, which we treat as an additional free parameter.\footnote{We will consistently assume that right-handed neutrinos, if present, are heavy enough so that they effectively decouple and do not play any relevant role. Models with light $\nu_R$ in connection to the $B$-anomalies can be found in~\cite{Iguro:2018qzf,Asadi:2018wea,Greljo:2018ogz,Robinson:2018gza}, and in connection to the vector leptoquark in~\cite{Azatov:2018kzb}.}

We define the general Lagrangian for these vectors as follows:
\begin{align}
\label{eq:U1Lag}
\mathcal{L}_{U_1}&=-\frac{1}{2}\,U_{1\,\mu\nu}^\dagger\, U_1^{\mu\nu}+M_U^2\,U_{1\,\mu}^\dagger\, U_1^{\mu}-ig_s\,(1-\kappa_U)\,U_{1\,\mu}^\dagger\,T^a\,U_{1\,\nu}\,G^{a\,\mu\nu}\nonumber\\
&\quad-ig_Y\,\frac{2}{3}\,(1-\tilde\kappa_U)\,U_{1\,\mu}^\dagger\,U_{1\,\nu}\,B^{\mu\nu}+\frac{g_U}{\sqrt{2}}\,[U_1^\mu\,(\beta_L^{ij}\,\bar q^i_L\gamma_\mu\ell_L^j+\beta_R^{ij}\,\bar d^i_R\gamma_\mu e_R^j)+{\rm h.c.}]\,,\\[5pt]
\label{eq:ZpLag}
\mathcal{L}_{Z^\prime}&=-\frac{1}{4}\,Z_{\mu\nu}^\prime\, Z^{\prime\,\mu\nu}+\frac{1}{2}\,M_{Z^\prime}^2\,Z_\mu^\prime\, Z^{\prime\,\mu}+\frac{g_{Z^\prime}}{2\sqrt{6}}\,Z^{\prime\,\mu}\,(\zeta_q^{ij}\,\bar q^i_L\,\gamma_\mu\, q_L^j+\zeta_u^{ij}\,\bar u^i_R\,\gamma_\mu\,u_R^j\nonumber\\
&\quad+\zeta_d^{ij}\,\bar d^i_R\,\gamma_\mu\, d_R^j-3\,\zeta_\ell^{ij}\,\bar \ell^i_L\,\gamma_\mu\, \ell_L^j-3\,\zeta_e^{ij}\,\bar e_R^i\,\gamma_\mu\, e_R^j)\,,\\[5pt]
\label{eq:GpLag}
\mathcal{L}_{G^\prime}&=-\frac{1}{4}\,G_{\mu\nu}^{\prime\,a}\, G^{\prime\,a\,\mu\nu}+\frac{1}{2}\,M_{G^\prime}^2\,G_\mu^{\prime\,a} \,G^{\prime\,a\,\mu}+\frac{1}{2}\,\kappa_{G^\prime}\,G_{\mu\nu}^{\prime\,a}\, G^{a\,\mu\nu}+g_s\,\tilde\kappa_{G^\prime} f_{abc}\,G_{\mu}^{\prime\,a}\,G_{\nu}^{\prime\,b}\, G^{c\,\mu\nu}\nonumber\\
&\quad+g_{G^\prime}\,G^{\prime\,a\,\mu}\,(\kappa_q^{ij}\,\bar q^i_L\,T^a\,\gamma_\mu\, q_L^j+\kappa_u^{ij}\,\bar u^i_R\,T^a\,\gamma_\mu\,u_R^j+\kappa_d^{ij}\,\bar d^i_R\,T^a\,\gamma_\mu\, d_R^j)\,,
\end{align}
where $T^a=\lambda^a/2$, with $\lambda^a$ ($a=1,\dots,8)$ the Gell-Mann matrices. 
In both $\mathcal L_{U_1}$ and $\mathcal L_{G'}$ we include possible non-minimal 
interactions with SM gauge fields, which play a role in the  pair  production of the heavy vectors at the LHC. 
In gauge models these couplings vanish, $\kappa_U=\tilde\kappa_U=\kappa_{G^\prime}=\tilde\kappa_{G^\prime}=0$.
However, this is not necessarily the case in strongly interacting models. 
The so-called minimal-coupling scenario for the leptoquark corresponds to $\kappa_U=\tilde\kappa_U=1$. 
Since a triple coupling of the type $GGG^\prime$ would lead to a huge enhancement 
of the colour-octet production at LHC and with that, to very strong constraints 
from high-energy data, in what follows we will take $\kappa_{G^\prime}=0$.

Without loss of generality, we choose the flavour basis of the $SU(2)_L$ fermion doublets to be aligned to the down-quark sector, i.e.
\begin{align}
q_L^i=
\begin{pmatrix}
V^*_{ji}\, u_L^j\\
d_L^i
\end{pmatrix}\,,
\qquad\qquad
\ell_L^i=
\begin{pmatrix}
\nu_L^i\\
e_L^i
\end{pmatrix}\,,
\end{align}
where $V_{ji}$ denote the CKM matrix elements.
We assume that the new vectors are coupled dominantly to third generation fermions.
The couplings to light quarks are assumed to
respect a $\mathrm{U(2)}_q$ flavour symmetry broken only 
in the leptoquark sector by the same leading 
spurion controlling the $3\to q$ mixing in the CKM matrix~\cite{Barbieri:2011ci}.
We parameterise the strength of this spurion by $\beta_L^{23}$. 
In the lepton sector we assume vanishing couplings to electrons. 
These assumptions are phenomenologically motivated by the tight constraints from low-energy 
observables, in particular $\Delta F=2$ amplitudes and lepton flavour violation in charged leptons 
(see e.g.~\cite{Buttazzo:2017ixm,Bordone:2018nbg}). 
More precisely, we take the following textures for the vector couplings ($Q=q,u,d$):
\begin{align}
\begin{aligned}
\beta_L&=\begin{pmatrix}0 & 0 & \beta_L^{13} \\ 0 & 0 & \beta_L^{23} \\0 & \beta_L^{32}  & \beta_L^{33}\end{pmatrix}\,,
&&&\beta_R&=\mathrm{diag}(0,0,\beta_R^{33})\,,\\[10pt]
\zeta_\ell&=\begin{pmatrix}0 & 0 & 0\\ 0 & \zeta_\ell^{22} & \zeta_\ell^{23}\\ 0 & (\zeta_\ell^{23})^* & \zeta_\ell^{33} \end{pmatrix}\,,&&&\zeta_e&=\mathrm{diag}(0,\zeta_e^{22},\zeta_e^{33})\,,\\[10pt]
\zeta_Q&=\mathrm{diag}(\zeta_Q^{ll},\zeta_Q^{ll},\zeta_Q^{33})\,,&&&\kappa_Q&=\mathrm{diag}(\kappa_Q^{ll},\kappa_Q^{ll},\kappa_Q^{33})\,.
\label{eq:lagrangian}
\end{aligned}
\end{align}
As shown in~\cite{Buttazzo:2017ixm},
the assumption of a single $\mathrm{U(2)}_q$ breaking spurion in both leptoquark and SM Yukawa couplings
implies the relation $\beta_L^{13}=V_{td}^*/V_{ts}^*\;\beta_L^{23}$. 
More generally, from $\mathrm{U(2)}$
symmetries acting on both quark and lepton sectors, we expect the following hierarchy:
 $|\beta_{L}^{3 1}|  \ll  |\beta_{L}^{2 3}|,~|\beta_{L}^{3 2}|  \ll |\beta_{R}^{3 3}|,~|\beta_{L}^{3 3}|= \mathcal{O}(1)$,
and analogously for the $\zeta_{\ell,e,Q}^{ij}$ and $\kappa_Q^{ij}$ couplings. 

Since our main motivation for analysing the high-$p_T$ phenomenology  of the $U_1$ is its 
success in addressing $B$-physics anomalies, it is worth recalling for which values of its 
couplings this can happen.  Detailed analyses of this question in specific UV models 
can be found for instance in~\cite{DiLuzio:2017vat, Bordone:2018nbg, DiLuzio:2018zxy, Cornella:2019hct}.
To remain sufficiently generic, we focus here on the  contribution to 
$R_{D}=\Gamma(B\to D \tau\nu)/\Gamma(B\to D \mu\nu)$. This  is the most interesting 
low-energy observable to constrain the $U_1$ couplings relevant at high-$p_T$ in the presence of 
right-handed currents.  Setting $\beta_{L}^{3 3}=-1$ one has~\cite{Cornella:2019hct}
\begin{align}\label{eq:RD}
\frac{ R_{D} - R_{D}^{\mathrm{SM}} }{  R_{D}^{\mathrm{SM}}  } 
\approx 0.2 \times \left(\frac{ g_U}{3 }\right)^2 \times \left[0.3 + 0.7 \left(\frac{\beta_{L}^{23}}{0.1}\right) \right]
\times \left\{  \ba{l l}   \left(\frac{ 2.2~{\rm TeV} }{M_U} \right)^2 \quad {\rm for}\quad  \beta_R^{33}=0 &  
\\[3pt]
  \left(\frac{ 4.2~{\rm TeV} }{M_U} \right)^2\quad {\rm for}\quad  \beta_R^{33}=-1  & 
 \ea
 \right.\,.
 \end{align}
This expression illustrates well the dependence on the relevant couplings. For $g_U$ within the pertubative regime
$(g_U < \sqrt{4\pi})$, the $U_1$ mass cannot exceed a few TeV. As  a reference benchmark, a $20\%$ enhancement in $R_{D}$ (in good agreement with present data) is
 obtained for $M_U=2.2$~TeV if   $\{g_U, \beta_{L}^{23} \} = \{3,0.1 \}$ and $\beta_{R}^{33}= 0$,
  or $M_U=4.2$~TeV with the same values of   $\{g_U, \beta_{L}^{23} \}$
but setting $\beta_{R}^{33}= -1$.

	\section{Results}
	\label{sec:results}	

We consider a variety of high-$p_T$ searches at the LHC which place limits on the model discussed above.  
The most constraining ones, which we discuss in detail below, are shown in \cref{tab:exp-searches}. 
In some cases the searches are optimised for the BSM processes we are interested in, 
allowing a simple translation of the reported limits in terms  model parameters.
In most cases however, a reinterpretation of the reported limits is necessary.

\begin{table}[t!]
\setlength{\tabcolsep}{10pt}
\renewcommand{\arraystretch}{1.5}
\centering
\begin{tabular}{lllcc}
\hline
Constrained BSM amplitude & Final state & Data set & Section & Reference\\
\hline
$U_1$ pair prod.       & $b\bar{b}\tau^-\tau^+$                & CMS, 35.9 fb$^{-1}$ & \ref{sec:pprod}     & \cite{Sirunyan:2018vhk}\\
$U_1$ pair prod.       & $t\bar{t}\nu_\tau \bar{\nu_\tau}$ & CMS, 35.9 fb$^{-1}$ & \ref{sec:pprod}     & \cite{Sirunyan:2018kzh}\\
$G'$ pair prod.       & $2b2\bar{b}$                & CMS, 35.9 fb$^{-1}$ & \ref{sec:pprod}     & \cite{Sirunyan:2018rlj}\\
$U_1$~[$t$ chan.] \&  $Z'$~[$s$ chan.]\ & $\tau_h^+ \tau_h^-$                    & ATLAS, 36.1 fb$^{-1}$ & \ref{sec:pp2tata}   & \cite{Aaboud:2017sjh}\\
$U_1$~[$t$ chan.]                & $\tau_h \nu$                               & CMS, 35.9 fb$^{-1}$ & \ref{sec:pp2tanu}  & \cite{Sirunyan:2018lbg}\\
$U_1$~[$t$ chan.] \&  $Z'$~[$s$ chan.]\ & $\tau_h \mu$                             & ATLAS, 36.1 fb$^{-1}$ & \ref{sec:pp2tamu} & \cite{Aaboud:2018jff}\\
$G'$~[$s$ chan.] \& $Z'$~[$s$ chan.]     & $t\bar{t}$                                    & ATLAS, 36.1 fb$^{-1}$ & \ref{sec:pp2tt}       & \cite{Aaboud:2018eqg}\\                           
\hline
\end{tabular}
\caption{Summary of the relevant experimental constraints.  All searches have a centre of mass energy of 13 TeV. }
\label{tab:exp-searches}
\end{table}

A relatively simple case is that of the leptoquark pair production. The differential and total cross-sections for these processes are well-known~\cite{Blumlein:1996qp}. Here we use the recent CMS analyses dedicated to 
pair-produced (scalar) leptoquarks decaying primarily to third generation SM fermions~\cite{Sirunyan:2018vhk,Sirunyan:2018kzh}.  
Since the leptoquarks are predominantly produced via their strong couplings to gluons, the limits only depend 
on the branching ratios to the relevant final states. Bounds on the coloron mass are extracted from a search 
for pair-produced resonances decaying to quark pairs, reported by CMS in the same way~\cite{Sirunyan:2018rlj}.

The case of the $\tau^+\tau^-$  final state, which constrains both the $Z^\prime$ 
($s$ channel production) as well as the $U_1$ ($t$ channel exchange), is significantly more involved. Here we re-interpret the limits 
on resonances decaying into tau-lepton pairs, with hadronically decaying taus, reported by ATLAS~\cite{Aaboud:2017sjh}\footnote{We do not consider the corresponding analysis by CMS~\cite{Sirunyan:2018zut}, which focuses on heavy Higgs bosons.}
(bounds from leptonic tau decays turn out to be significantly weaker at large ditau invariant masses).
We first consider the bounds placed on the $U_1$ and on the $Z^\prime$  
in isolation, for various choices of couplings and widths, and then in combination.  
As we emphasise below, it is essential to include all relevant experimental information when deriving limits in this case.

We extract further bounds on $U_1$ by recasting CMS searches for $pp\to\tau \nu$~\cite{Sirunyan:2018lbg} and limits on both $Z^\prime$ and $U_1$ from the $pp\to \tau \mu$ search by ATLAS~\cite{Aaboud:2018jff}.
In both cases the 2-3 family mixing of the leptoquark plays a key role. As far as other dilepton final states are concerned,
we explicitly checked that constraints from $pp\to\mu\mu$ (see e.g.~\cite{Aaboud:2017buh}) do not significantly constrain
the parameter space relevant to our model.

The leading bound on the $G^\prime$ is extracted by the unfolded
$t\bar{t}$ invariant mass spectrum provided by ATLAS~\cite{Aaboud:2018eqg}.
In principle, the $U_1$ and the $Z'$ could be constrained by dijet searches. However, in our setup resonances 
tend to be very wide, with a width-over-mass $\sim 25\%$. 
As a result, the limits reported in the literature on narrow dijet peaks over a data driven 
background spectrum~\cite{Sirunyan:2018xlo,CMS:2018wxx,Aaboud:2018fzt} are not directly applicable. 
Furthermore, dijet signatures are mostly produced for light quarks and gluons, which couple only weakly to $Z'$ and $G^\prime$ in our 
setup.\footnote{Searches for $b$-tagged dijet signatures would remedy this, but they tend to be sensitive to relatively low mass ranges and narrow widths~\cite{Sirunyan:2018pas,Aaboud:2018tqo}.} 
Indeed, dedicated recasts of dijet searches performed in a setup similar to ours 
have shown that these constraints are less significant than those from the $t\bar{t}$ final state~\cite{DiLuzio:2018zxy}. 
Although one can envision scenarios where current dijet searches are more constraining than $t\bar t$ searches, such as when third-generation couplings are suppressed or when light-generation couplings are large, these limits are less relevant for the class of models which fit the flavour anomalies and so we do not consider dijet searches.

To perform recasts of these searches we implement the model described in~\cref{sec:model} in {\tt FeynRules 2.3.32}~\cite{Alloul:2013bka} and generate the corresponding UFO model file. The FeynRules model files as well as the corresponding UFO model are available at \href{https://feynrules.irmp.ucl.ac.be/wiki/LeptoQuark}{\tt https://feynrules.irmp.ucl.ac.be/wiki/LeptoQuark}. In our Feynrules implementation and in all our results throughout this paper, we include only tree-level effects. While some NLO QCD corrections are available for the vector leptoquark case~\cite{Aebischer:2018acj}, in specific models these are expected to be supplemented by additional NLO contributions that can be of similar (or even larger) size. Hence we opt not to include them and we add a systematic error in our signal to (partially) account for them. Other Feynrules implementations for the vector leptoquark (but with interactions to third-generation left-handed fields only) are available~\cite{Dorsner:2018ynv}. We have cross-checked our leptoquark implementation (with $\beta_R^{33}=0$) against the one in~\cite{Dorsner:2018ynv}, finding a perfect agreement between the two.

\subsection{Limits from resonance pair production}
\label{sec:pprod}	
	
We first briefly discuss limits on the leptoquark coming from their pair production. For a large fraction of the parameter space, the dominant production modes are governed by QCD and the relevant couplings are the strong gauge coupling and $\kappa_U$, see \cref{eq:U1Lag}. However, for $\kappa_U\sim1$ the QCD-induced production cross-section is smaller and pair-production via lepton exchange becomes relevant for large values of $g_U$.  The most constraining searches in our scenario are those for the $b\bar{b}\tau^-\tau^+$ \cite{Sirunyan:2018vhk} and $t\bar{t}\nu_\tau \bar{\nu_\tau}$ \cite{Sirunyan:2018kzh} final states.

\begin{table}[t!]
\setlength{\tabcolsep}{10pt}
\renewcommand{\arraystretch}{1.5}
\centering
\begin{tabular}{cccccccc}
\hline
 \multicolumn{2}{c}{Parameters} && 
 \multicolumn{2}{c}{$b\bar{b}\tau^-\tau^+$ final state} && 
 \multicolumn{2}{c}{$t\bar{t}\nu_\tau \bar{\nu_\tau}$ final state}\\
 $\kappa_U$ & $\beta_R^{33}$ && 
 $\text{BR}(U_1\to b \tau^+)$& Limit [TeV]&&
 $\text{BR}(U_1\to t \bar{\nu_\tau})$& Limit [TeV]\\
  \hline
0&0 && 0.51 & 1.4 && 0.50 & 1.6\\
0&1 && 0.67 & 1.5 && 0.33 & 1.3\\
1&0 && 0.51 & $1.1-1.3$ && 0.49 & $1.1-1.2$\\
1&1 && 0.67 & $1.2-1.4$ && 0.32 & $1.0-1.2$\\
\hline
\end{tabular}
\caption{Summary of the experimental constraints on pair produced leptoquarks in the $b\bar{b}\tau^-\tau^+$ \cite{Sirunyan:2018vhk} and $t\bar{t}\nu_\tau \bar{\nu_\tau}$ \cite{Sirunyan:2018kzh} final states, assuming the leptoquarks decay solely into third generation SM particles. When $\kappa_U = 1$, QCD production processes become less important and lepton exchange (which depends on $g_U$) is relevant.  We thus show how the limit varies in the range $g_U \in [0,4]$.}
\label{tab:lq-pair-production}
\end{table}

In \cref{tab:lq-pair-production} we report the limits for various values of $\kappa_U$ and $\beta_R^{33}$, which determines the branching ratios (the branching ratios deviate slightly from the expected 1/2, 1/3, 2/3 due to phase space effects).  We assume that the leptoquark decays only into third generation SM particles and  find that the limits range from 1 TeV to 1.6 TeV. Similar limits have also been obtained in the literature, see e.g.~\cite{Sirunyan:2018kzh,Angelescu:2018tyl,Diaz:2017lit}, although using lower luminosity in the $b\bar{b}\tau^-\tau^+$ channel. Whenever it is possible to compare, we find good agreement between our results and those in the aforementioned references. With $\kappa_U=0$ there is an extra coupling to the gluon field strength tensor boosting the production cross-section and strengthening the limit.  As $\beta_R^{33}$ increases, the branching ratio to $ b \tau^+$ increases while the branching ratio to $t \bar{\nu}_\tau$ decreases, which is reflected in a strengthening and weakening of the limits, respectively.  For illustration, we include the strongest bound from pair-production, i.e the limit $M_U>1.6$~TeV, in \cref{fig:LQ_excl,fig:LQ_tanu_excl,fig:tamu_excl}.

In a similar fashion, bounds on the coloron mass can be extracted from a search for pair-produced resonances decaying to quark pairs, performed by the CMS collaboration~\cite{Sirunyan:2018rlj}. The search excludes a coloron in the whole mass range considered, from 80~GeV to 1500~GeV, so provides an upper bound of $M_{G'}>1.5$~TeV.  However, a stronger upper bound can be estimated by extrapolating the production cross-section and exclusion limit to higher energies, where bounds of 1.7~TeV and~2.1~TeV for~$\tilde\kappa_{G'}=0$ and~$\tilde\kappa_{G'}=1$ are obtained.  The stronger bound in the latter case can be understood from the fact that the corresponding operator in \cref{eq:GpLag} adds significantly to the $gg\to G' G'$ amplitude.    The estimated limits are practically independent of the choices of the couplings to quarks, because the production cross section is dominated by the gluon-initiated processes. In setting these limits, we fix the coloron gauge coupling to $g_{G'}=3$, $\kappa_{G'}=0$ and $\kappa_{q,u,d}^{33}=1$.

\subsection{\texorpdfstring{$pp\to\tau\tau$}{Ditau} search}
\label{sec:pp2tata}	

The ATLAS collaboration has performed a search of heavy resonances in the ditau final state using $36.1~\textrm{fb}^{-1}$ of 13~TeV data~\cite{Aaboud:2017sjh}. In this section we recast this search to set limits on the $U_1$ and $Z^\prime$ masses for different choices of the couplings. In \cref{sec:pp2tata_Zp} and \cref{sec:pp2tata_U1} we consider separate limits for the $Z^\prime$ and $U_1$ assuming that one of the two has fully decoupled. The interplay of the two resonances in this search is considered at the end, in \cref{sec:pp2tata_allin}.
	
\begin{table}[t!]
\setlength{\tabcolsep}{20pt}
\renewcommand{\arraystretch}{1.5}
\centering
\begin{tabular}{cc}
\hline
Particle selection & At least two $\tau_h$'s and no electrons or muons\\
Charge & $\tau_{h1}$ $\tau_{h2}$ should be of opposite charge\\
$\tau_h$ $p_T$ & $p_T^{\tau_{h1}}>130$~GeV, $p_T^{\tau_{h2}}>65$~GeV\\
$\eta$ & $|\eta_{\tau_h}|<2.5$ excluding $1.37<|\eta_{\tau_h}|<1.52$\\
$\phi$ & $|\Delta\phi(\tau_{h1},\tau_{h2})|>2.7$~rad\\
\hline
\end{tabular}
\caption{Summary of the experimental cuts for the ATLAS $\tau_h\tau_h$ search~\cite{Aaboud:2017sjh}. For the leading $\tau_h$ we use the $p_T$ cut $p_T^{\tau_{h1}}>130$~GeV as quoted in the HEPData entry for Ref.~\cite{Aaboud:2017sjh}. Note that the corresponding cut was $p_T^{\tau_{h1}}>85$~GeV for $10\%$ of the data.}
\label{tab:cutsDYtataATLAS}
\end{table}

\subsubsection{Search strategy}
\label{sec:pp2tata_strategy}	
We focus on the analysis with $\tau_h\tau_h$ since this channel presents the highest sensitivity to high-mass resonances. The contributions to the $pp\to\tau^+\tau^-$ process from new heavy resonances, including the interference with the SM, are computed using {\tt Madgraph5\_aMC@NLO v2.6.3.2}~\cite{Alwall:2014hca}, with the {\tt NNPDF23\_lo\_as\_0119\_qed} PDF set~\cite{Ball:2012cx}. Hadronization of the $\tau$ final states is performed with {\tt Pythia 8.2}~\cite{Sjostrand:2014zea} with the A14 set of tuned parameters~\cite{ATL-PHYS-PUB-2014-021}. Detector simulation is done using {\tt Delphes~3.4.1}~\cite{deFavereau:2013fsa}. The ATLAS Delphes card has been modified to satisfy the object reconstruction and identification requirements. In particular we include the $\tau$-tagging efficiencies quoted in the experimental search~\cite{Aaboud:2017sjh}. After showering and detector simulation, we apply selection cuts using {\tt MadAnalysis~5 v1.6.33}~\cite{Dumont:2014tja} (see \cref{tab:cutsDYtataATLAS} for details on the applied cuts). We have validated our results by generating the SM Drell-Yan $pp\to\tau\tau$ background and comparing our results with the one quoted by ATLAS. A good agreement is found between the two samples (we find a discrepancy with the quoted central values of less than 20\%, well within the given $1\sigma$ region).

After passing through selection cuts, the resulting events are binned according to their total transverse mass,
\begin{align}
m_T^{\rm tot}\equiv\sqrt{(p_T^{\tau_{h1}}+p_T^{\tau_{h2}}+E_T^{\rm\,miss})^2-(\vec{p}_T^{\;\tau_{h1}}+\vec{p}_T^{\;\tau_{h2}}+ \vec{p}_T^{\;\rm miss} )^2}\,,
\end{align}
where $p_T^{\tau_{h1,2}}$ are the transverse momenta of the visible decay products for the leading and sub-leading taus, respectively, and 
$E_T^{\rm\, miss}$ and $\vec{p}_T^{\;\rm miss}$ are the total missing transverse energy and missing momentum in the reconstructed event. We compare our binned events with the histogram in Fig.~3b of the supplementary material of~\cite{Aaboud:2017sjh}, which contains the corresponding $m_T^{\rm tot}$ histograms for the SM background and the experimental data, with $b$-tag inclusive event selection. For the statistical analysis we use the modified frequentist $\textrm{CL}_s$ method~\cite{Read:2002hq}. We compute the $\textrm{CL}_s$ using the {\tt ROOT}~\cite{Brun:1997pa} package {\tt Tlimit}~\cite{Junk:1999kv} and exclude model parameter values with $\textrm{CL}_s<0.05$. In our statistical analysis we include all the bins and SM backgrounds errors, provided by the ATLAS collaboration in the corresponding HEPData entry~\cite{Aaboud:2017sjh}.\footnote{The power of each of the bins in excluding a signal is shown in \cref{fig:impact-of-n-bins}, where we plot the 95\%~CL exclusion limit in the leptoquark mass, as a function of the number of the bins included in the statistical analysis.} When combining the bins, we ignore any possible correlation in the bin errors, since they are not provided by the experimental collaboration. We also include a systematic uncertainty of $20\%$ for the signal to account for possible uncertainties related to the PDF, tau hadronization, detector simulation and unaccounted NLO corrections. 

\begin{figure}[t!]
  \begin{center}
    \begin{tabular}{cc}
      \includegraphics[height=0.45\textwidth]{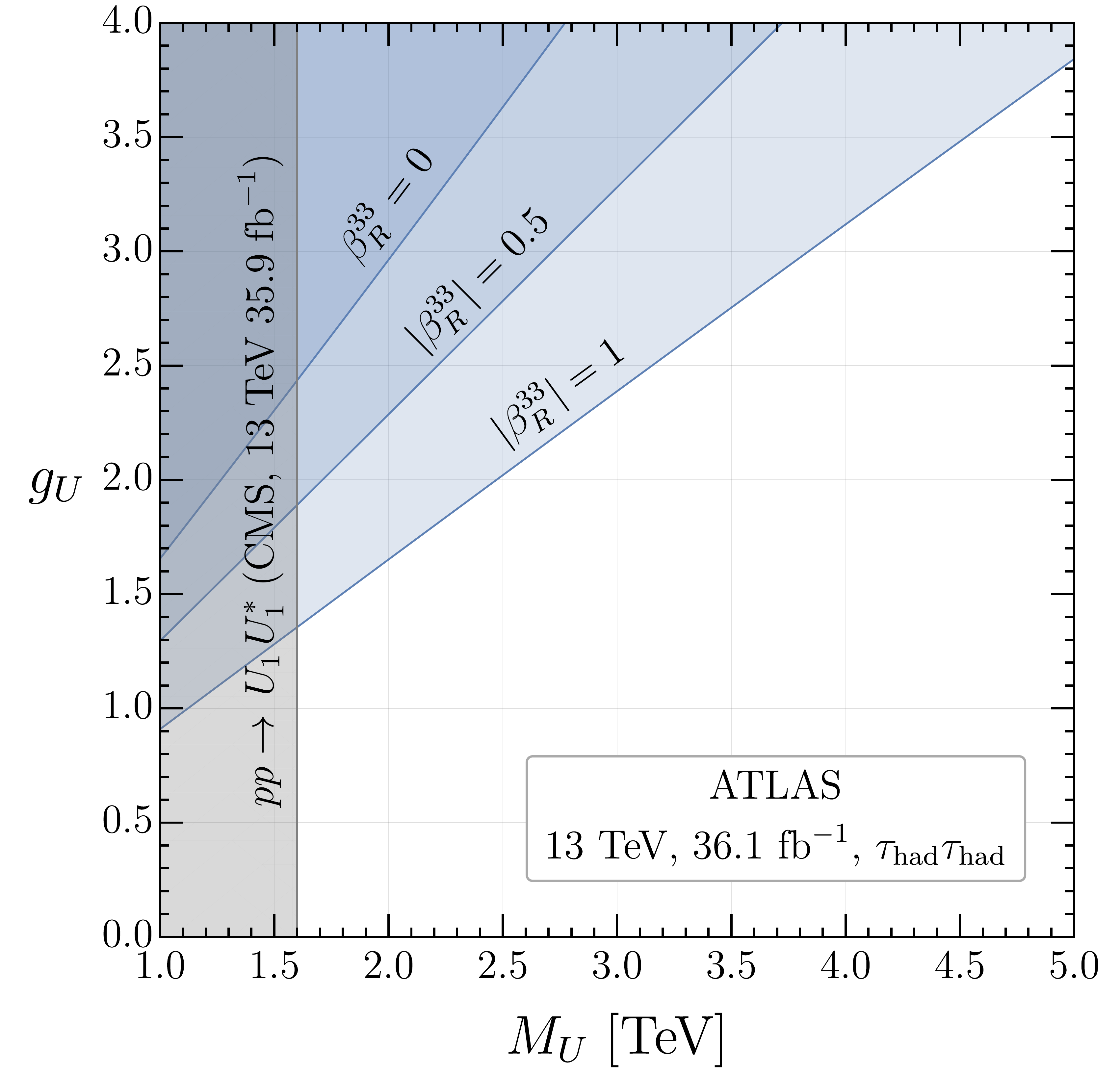} &   
      \includegraphics[height=0.45\textwidth]{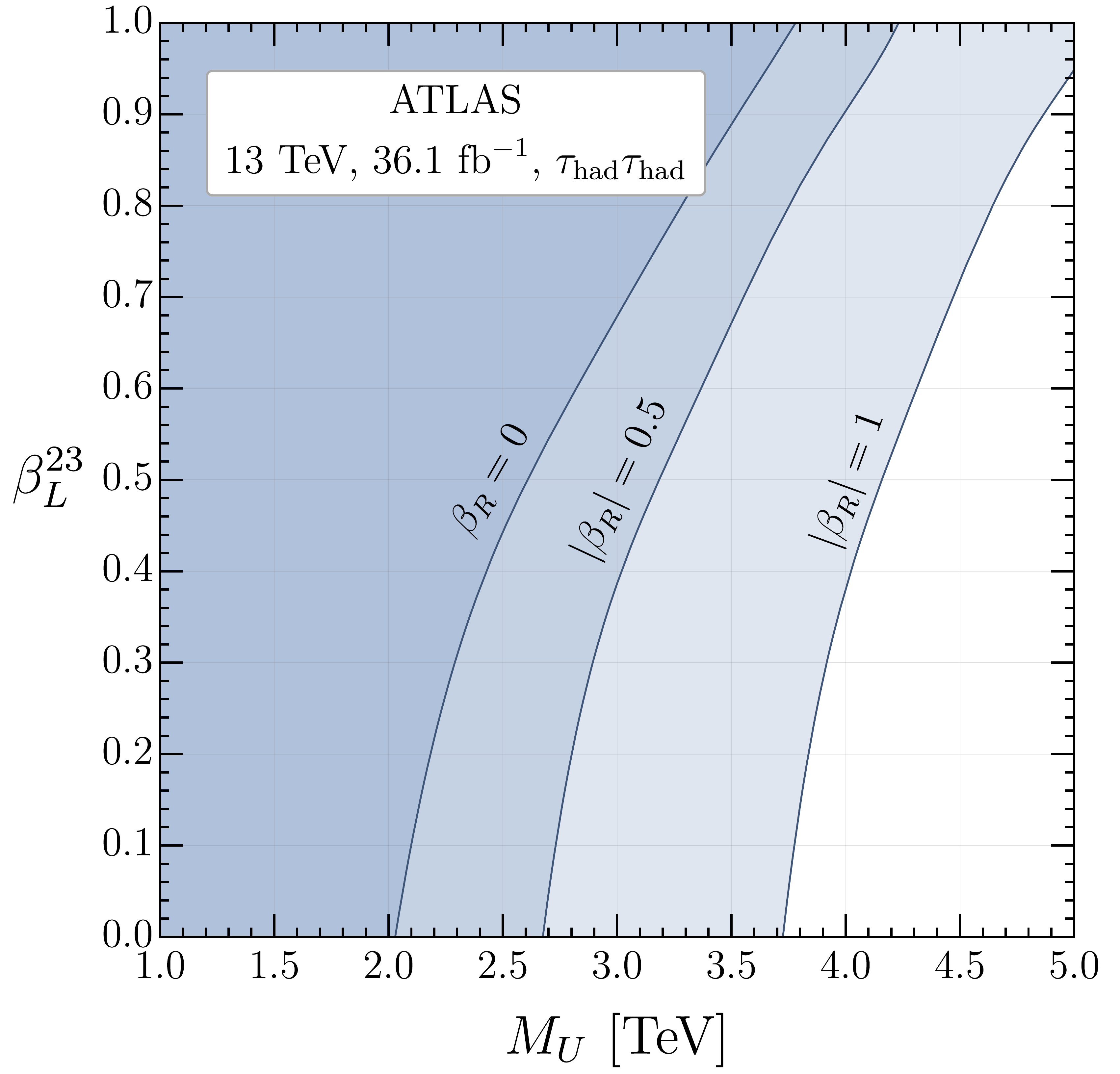}
    \end{tabular}
  \end{center}
\caption{Exclusion plot for the $pp\to\tau\tau$ search in the $(g_U,M_U)$ (left) and $(\beta_L^{23},M_U)$ (right) planes for different values of the  coupling $\beta_R^{33}$. We fix $\beta_L^{33}=1$ and the leptoquark width to its natural value. 
In the left plot we set $\beta_L^{23}=0$ and, for comparison, 
we also show the limits from $U_1$ pair production. In the right plot we set $g_U=3$.}
\label{fig:LQ_excl}
\end{figure}	

\subsubsection{Limits on the \texorpdfstring{$U_1$ leptoquark}{leptoquark}}
\label{sec:pp2tata_U1}	
In this section we decouple the $Z^\prime$ and concentrate on the limits arising exclusively from the leptoquark exchange. In our search we take maximal values for $\beta_L^{33}$ (i.e. $\beta_L^{33}=1$) and consider three benchmarks for the right-handed coupling: $|\beta_R^{33}|=\{0.0,\,0.5,\,1.0\}$. Note that the search is not sensitive to the relative sign choice between $\beta_R^{33}$ and $\beta_L^{33}$ but only to their magnitudes. The reason for this is that the New Physics (NP) amplitudes of different chiralities do not interfere with each other and the amplitude proportional to $\beta_R^{33}\, \beta_L^{33}$ does not interfere with the SM ones. We further fix the leptoquark width to its natural value. 
The leptoquark width only mildly
affects the results of this search, contrary to the $Z^\prime$ case discussed in the next section, since the NP contribution is generated via a $t$ channel exchange.

Exclusion limits in the $(g_U, M_U)$ plane, setting $\beta^{23}_L=0$, are shown in~\cref{fig:LQ_excl} (left). Similar recasts for the case with $\beta_R^{33}=0$ can be found in the literature~\cite{Angelescu:2018tyl,Schmaltz:2018nls}.  We obtain slightly stronger limits than those in the previous references. As we show in \cref{fig:impact-of-n-bins}, this difference can be understood from the fact that we consider the full $m_T^{\rm tot}$ distribution and not only the highest bin. The lower bins are important since a $t$ channel exchange gives rise to a broad tail in the spectrum. Exclusion limits for the scenario where $\beta_R^{33}\neq0$ have not been discussed in the literature. We find that the additional chirality significantly enhances the cross section, yielding limits that are about $70\%$ stronger than in the case when $\beta_R^{33}=0$.

Finally, we also study the limits on $M_U$ for non-zero values of $\beta_L^{23}$, \cref{fig:LQ_excl} (right). Here we fix $g_U=3$ and $\beta_L^{33}=1$ and plot the corresponding exclusion limits for the three benchmark values of $\beta_R^{33}$ discussed above. As can be seen, only a mild increase of the limits is found for $\beta_L^{23}\lesssim 0.4$. For larger values of $\beta_L^{23}$, the PDF enhancement is enough to make $s\bar s\to \tau^+\tau^-$ the dominant partonic channel and the limits start growing linearly with $\beta_L^{23}$.

\subsubsection{Limits on the \texorpdfstring{$Z^\prime$}{Z'} resonance}
\label{sec:pp2tata_Zp}	

We now proceed to the limits set on the $Z'$, decoupling the leptoquark. Throughout this section we fix $\zeta_{q,u,d}^{33} = \zeta_{\ell,e}^{33} =1$ and focus on the impact of varying the overall $Z'$ coupling $g_{Z'}$, varying the coupling to left-handed light quarks $\zeta_q^{ll}$, and varying the width of the $Z'$.

\begin{figure}[t!]
\centering
    \begin{tabular}{cc}
      \includegraphics[height=0.45\textwidth]{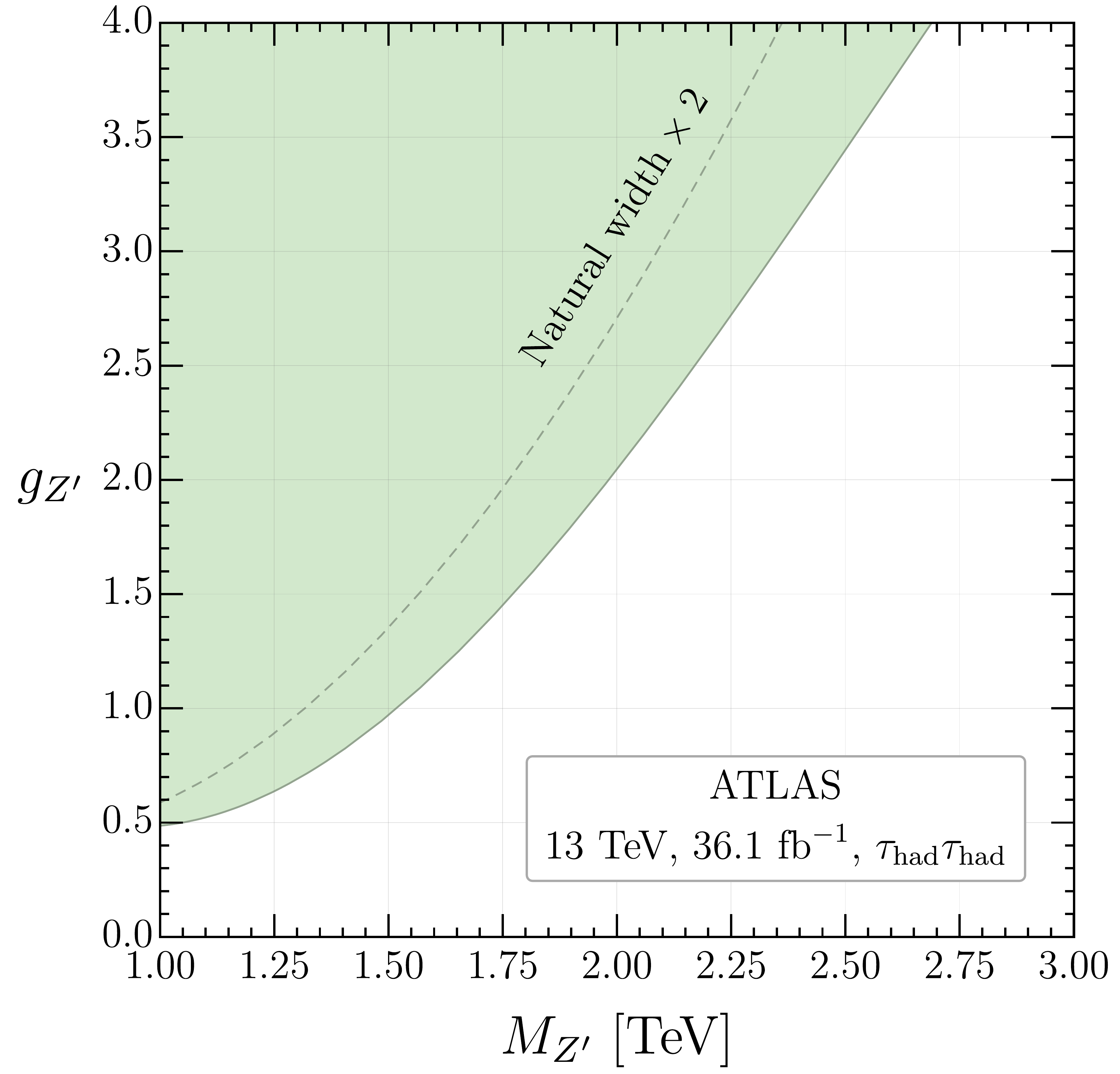} &
      \includegraphics[height=0.45\textwidth]{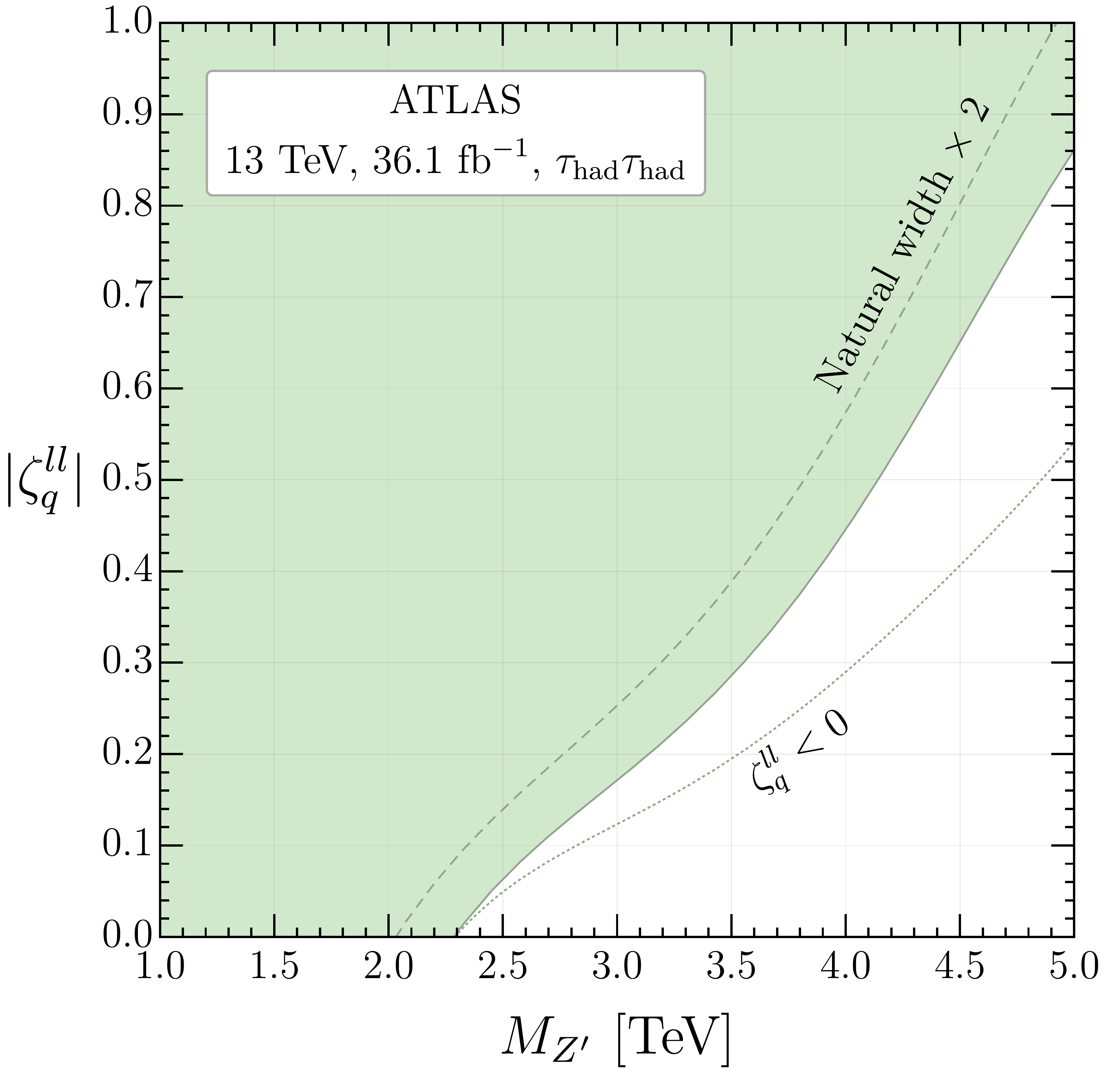}
    \end{tabular}
\caption{ Exclusion plot for the $pp\to\tau\tau$ search in the $(g_{Z'},M_{Z'})$ plane  (left) 
and $(\zeta_q^{ll},M_{Z'})$ plane (right),  and for the natural width $\times$ 2 while maintaining 
the natural partial width to tau pairs (dashed curves). 
In the left plot we set $\zeta_q^{ll}=0$.  In the right plot we set $g_{Z'} = 3$.}  
\label{fig:Zp_excl}
\end{figure}

In the left panel of \cref{fig:Zp_excl} we set $\zeta_q^{ll} = 0$ and show the exclusion in the $(g_{Z'},M_{Z'})$ plane.  
For small couplings,  $g_{Z'} < 0.5$, the $Z'$ is not excluded above 1 TeV as the production cross section is too small.  
In the range $0.5 < g_{Z'} < 1.0$ the limit increases from 1 TeV to 2 TeV and it approaches a regime where 
it increases linearly with the coupling.  This can be understood by the fact that, having set  $\zeta_q^{ll}=0$,
the $Z'$ is dominantly produced from $b$-quarks, which  carry only low momentum fractions of the protons. 
As a result, even for relatively low masses the effective cross-section scales like a contact 
interaction $\sigma_Z' \sim g_Z'^4/M_Z'^4$.

Finally, we also show the impact of varying the width. As can be noted, doubling the width (dashed line in \cref{fig:Zp_excl})
has a relatively minor impact. This is consistent with the observation that the limits does not come from 
the on-shell production of the $Z^\prime$,  but rather from its tail (that scales like a contact interaction).

In \cref{fig:Zp_excl} (right) we fix $g_{Z'} = 3$ and vary the couplings to left-handed light quarks $\zeta_q^{ll}$.  Since the light quarks have less PDF suppression than the third-generation quarks, the limit increases rapidly. For $\zeta_q^{ll} \ll 1$, the width is not affected by increasing $\zeta_q^{ll}$, while for larger values of  $\zeta_q^{ll}$ the width starts to be affected leading to a change of slope. 

We again show that doubling the natural width decreases the limit by around 10 \%.  We also show the impact of changing the relative sign between the light quark couplings and the third-generation coupling.  With opposite signs the interference term contributes constructively, strengthening the limit, whereas when the signs are the same the interference term contributes destructively, weakening the limit.

\begin{figure}[t!]
\centering
      \includegraphics[height=0.45\textwidth]{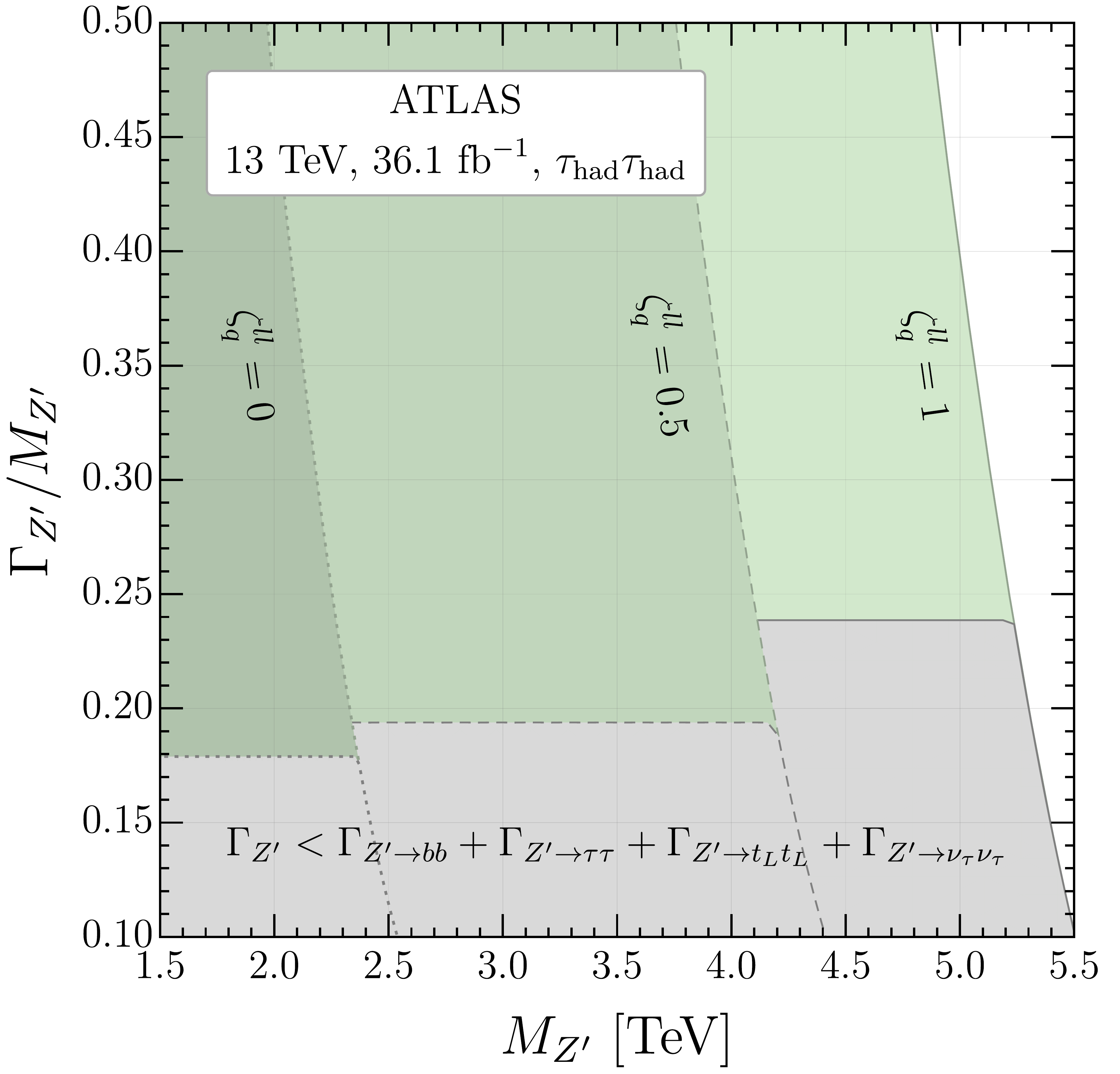} 
  \caption{Exclusion plot for the $pp\to\tau\tau$ search in the $(\Gamma_{Z'}/M_{Z'},M_{Z'})$ plane for $\zeta_q^{ll}=\{0.0,\, 0.5,\, 1.0\}$ for dotted, dashed and full lines, respectively. The grey area shows values of the width which are below the corresponding natural width.}
  \label{fig:Zp_width}
\end{figure}

In \cref{fig:Zp_width} we fix $g_{Z'} = 3$ and vary the width for $\zeta_q^{ll}\in\{0.0,\,0.5,\,1.0\}$.  As noted above, we see that the limit depends only weakly on the width.  For all values of $\zeta_q^{ll}$, a doubling of the width from 25\% to 50\% decreases the limit by around 10\%.  The grey area show values of the width which are below the corresponding natural width.


In summary, the $Z'$ mass limit of the ditau search 
depends weakly on the universal coupling $g_{Z'}$, is very sensitive to the light-quark couplings (it is excluded below 5 TeV for $\zeta_q^{ll} \approx 1$), and is only weakly relaxed by an increase of the total width of the $Z'$.

\subsubsection{Combined limits for the \texorpdfstring{$Z^\prime$}{Z'} and the \texorpdfstring{$U_1$ leptoquark}{leptoquark}}
\label{sec:pp2tata_allin}	

\begin{figure}[t!]
\centering
\includegraphics[width=0.5\textwidth]{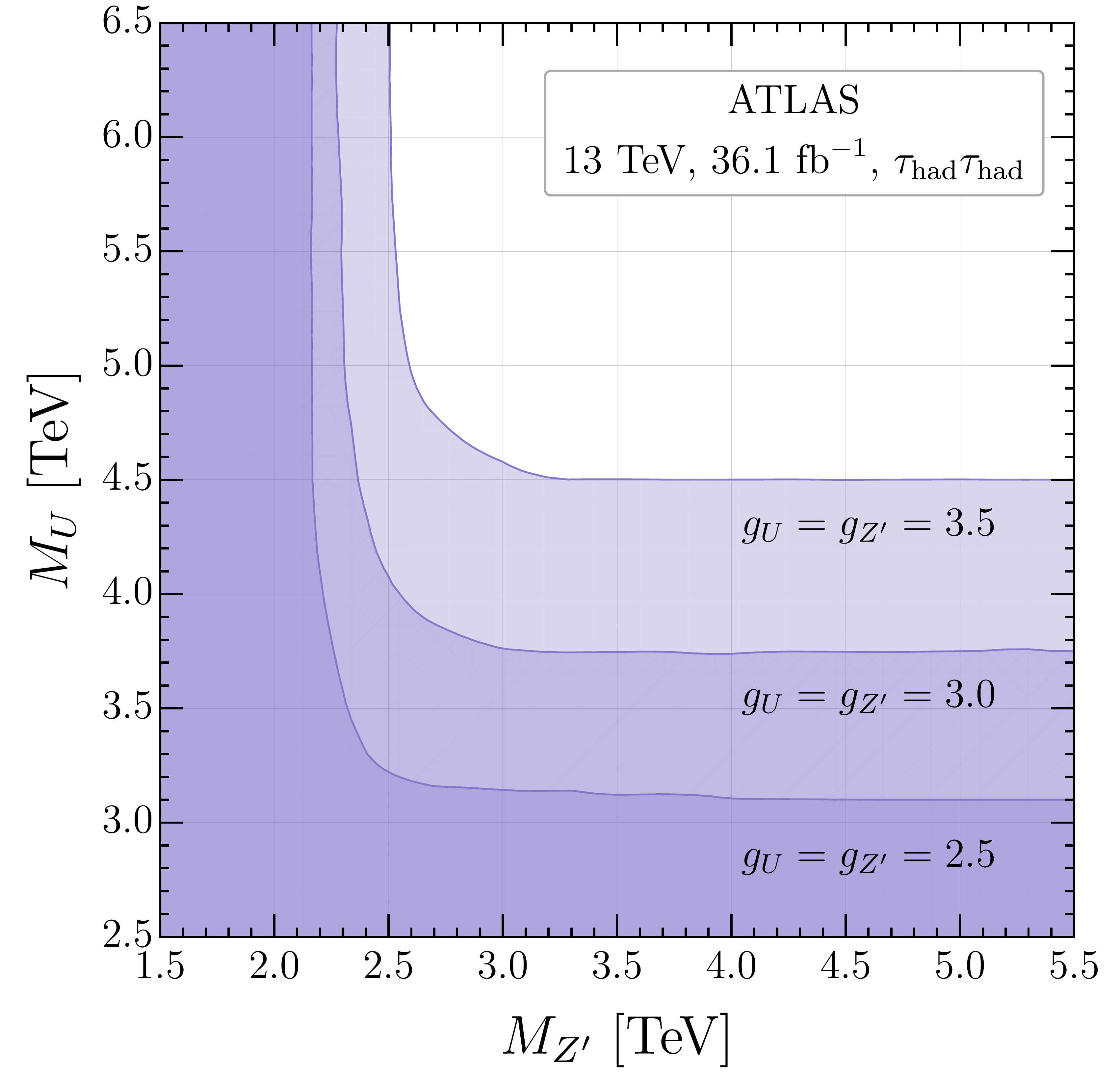} 
\caption{$95\%$ CL exclusion limits from the $pp\to\tau\tau$ search in the $(M_U,M_{Z^\prime})$ plane for different values of the gauge couplings $g_U=g_{Z^\prime}$.  All couplings to light quarks are set to zero.}
\label{fig:MZpLQ_excl}
\end{figure}	

We now consider the limits when both the $Z'$ and the leptoquark are present. 
For the $Z'$ we set $\zeta_{q,u,d}^{33} = \zeta_{\ell,e}^{33} =1$ and $\zeta_q^{ll} = 0$.  For the leptoquark we set $\beta_L^{33}=\beta_R^{33}=1$ and $\beta_L^{23}=0$. In both cases we assume natural widths.  

In \cref{fig:MZpLQ_excl} we show the exclusion limit on the $(M_U,~M_{Z'})$ plane for a variety of overall coupling strengths, $g_U = g_{Z'} \in \{2.5,\, 3.0,\, 3.5\}$.  The increase of the limits with growing coupling in each step is relatively small for the $Z'$ ($\sim 200 \GeV$), while it is larger for the leptoquark ($\sim 600 \GeV$). We see that the decoupling regimes considered in the previous two sections hold when the $Z'$ is heavier than (roughly) $3$~TeV, and when the leptoquark is heavier than $5-6$~TeV.

Below the decoupling regime, the limits on both particles strengthen by a few hundred GeV, since they both contribute to the $\mttot$ distribution.


\begin{figure}
  \begin{center}
    \begin{tabular}{cc}
      \includegraphics[height=0.43\textwidth]{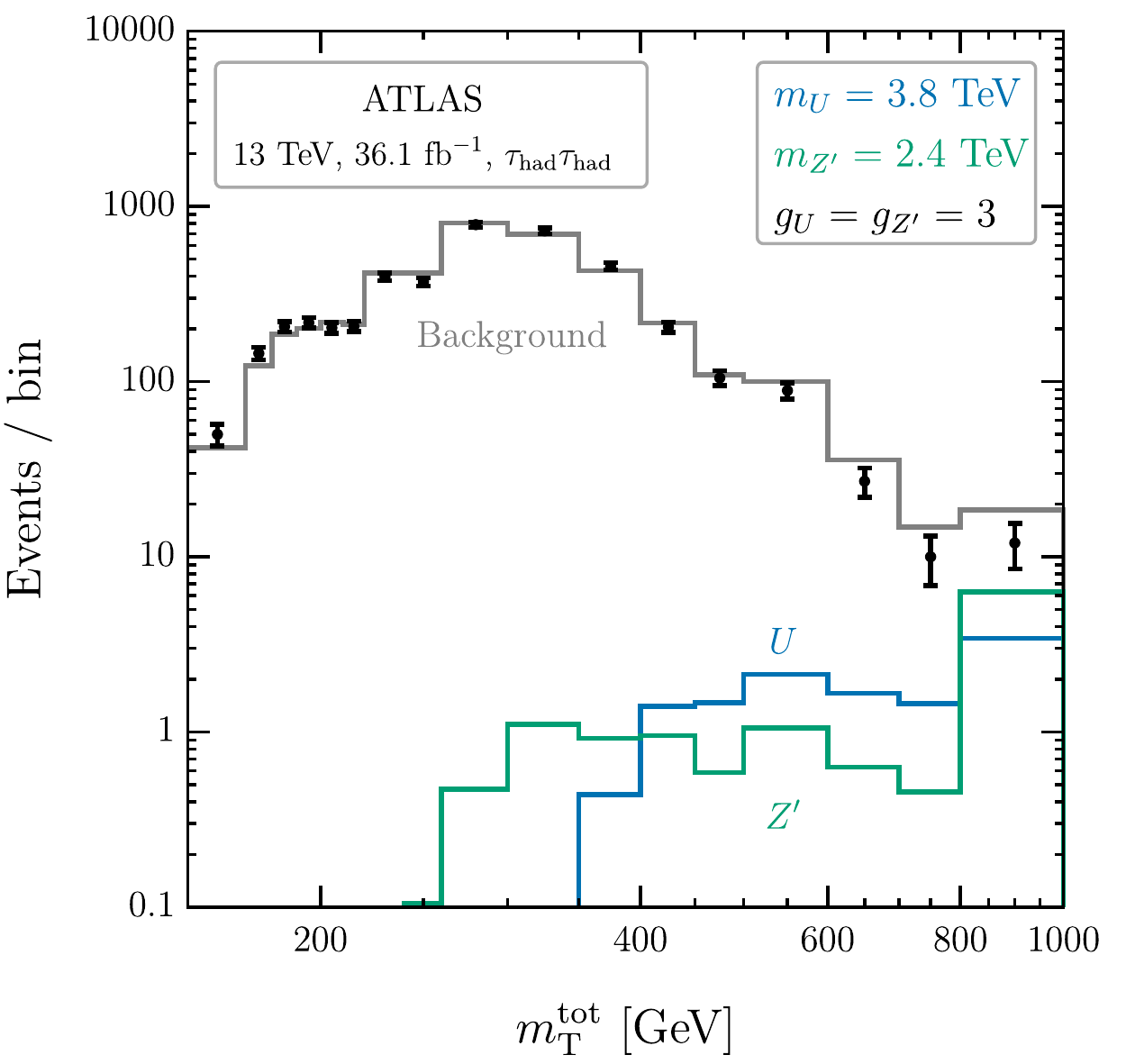} & 
      \includegraphics[height=0.43\textwidth]{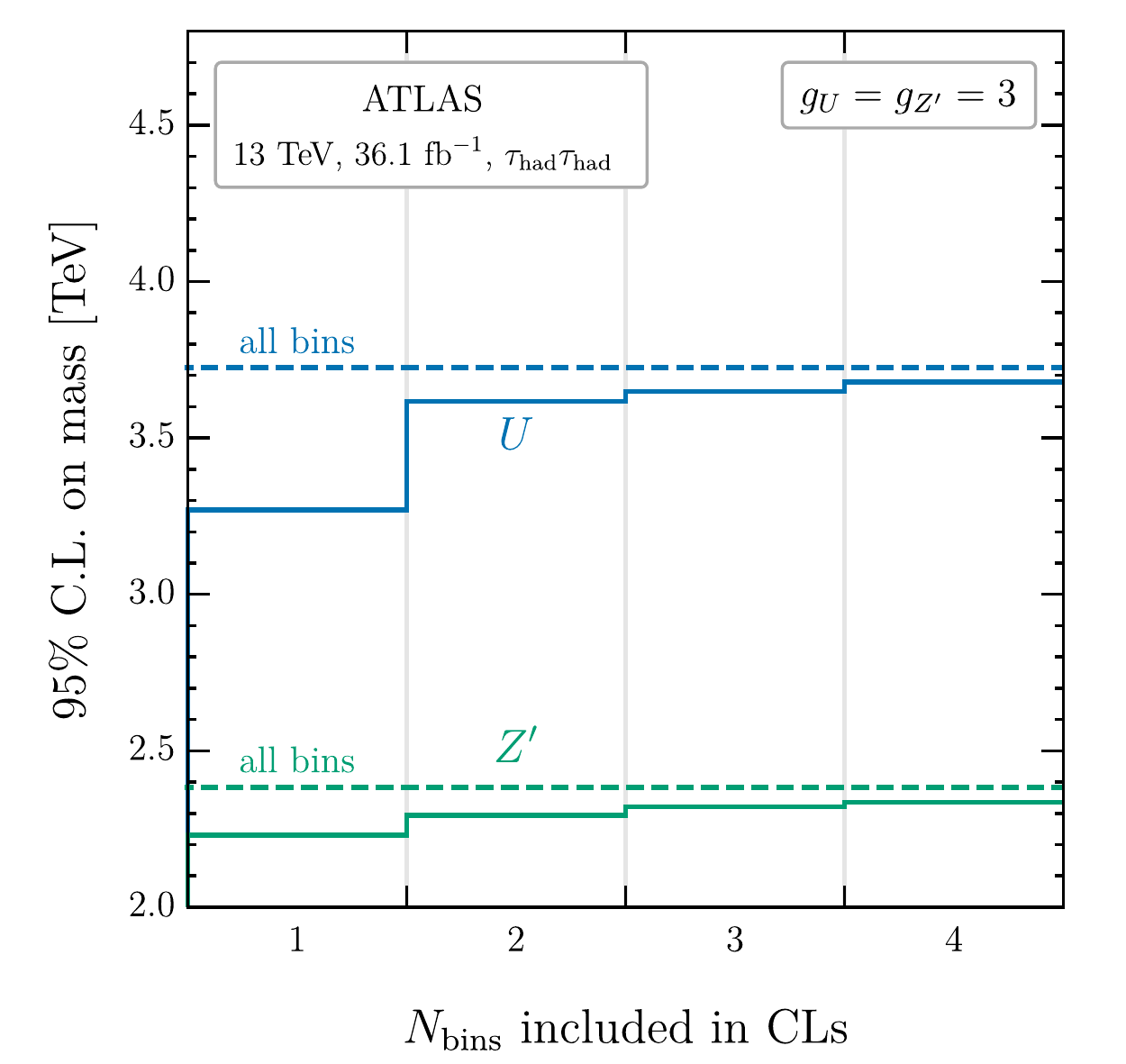}
    \end{tabular}
  \end{center}
  \caption{Left: Distributions in $\mttot$ for the $b$--inclusive background and data from~\cite{Aaboud:2017sjh}, 
  a leptoquark (blue) and a $Z'$ (green) with masses and couplings as shown.  
  Right:~The 95\%~C.L. on the mass of a leptoquark (blue) and a $Z'$ (green) 
  when only the highest $N$ bins are included in the CLs, with couplings as shown. 
  The dashed lines show the limit obtained when all bins are included.
   }
  \label{fig:impact-of-n-bins}
\end{figure}


\bigskip
We now highlight the importance of including more than just the highest bin in $\mttot$  in setting the mass limit.  
In \cref{fig:impact-of-n-bins} (left) we plot the $\mttot$ distribution of the data and background 
from~\cite{Aaboud:2017sjh}, along with our simulated leptoquark and $Z'$ contributions.  We show the distributions 
for $g_{U} = 3$ and $g_{Z'} = 3$, for masses at the 95\%~C.L. limit.  
After a peak, the background steadily falls with increasing $\mttot$.  The final bin has a larger number 
of events than the preceding bin as this bin is wider and as it includes overflow events.  
As such, the final three bins each contain a similar number of background events.  
Since tau pair production via a $Z'$ proceeds through an $s$ channel, 
it is more peaked in $\mttot$ and the events from a multi-TeV $Z'$ cluster in the highest energy bin.
However, tau pair production via a leptoquark proceeds through $t$ channel process, 
so there is no clear peak in the invariant mass distribution.  
This leads the distribution in $\mttot$ to extend to lower values.
We see in \cref{fig:impact-of-n-bins} (right) the impact of including only the $N$ highest bins in the CLs calculation.  
For the $Z'$, the limit obtained with only the highest bin is almost $200\GeV$ lower than the limit including all bins.
For the leptoquark, when only the highest bin is included, the 95\% C.L.~limit is around $400\GeV$ 
weaker than when all bins are included.  When the highest two bins are included the difference reduces to 
around $100\GeV$, and slowly improves as more bins are added.  
We see that it is crucial to include more than the highest bin in $\mttot$ to produce an accurate estimate 
of the leptoquark exclusion limit. However, it should be noted that we have not been able to account for 
possible correlations between the bin errors, which could impact the derived exclusion limits.

\subsection{\texorpdfstring{$pp\to\tau\nu$}{Tau + neutrino} search}
\label{sec:pp2tanu}	

The ATLAS and CMS collaborations have performed searches for heavy resonances decaying to $\tau\nu$ (with hadronically decaying $\tau$) using $36.1~\textrm{fb}^{-1}$~\cite{Aaboud:2018vgh} and $35.9~\textrm{fb}^{-1}$~\cite{Sirunyan:2018lbg} of 13~TeV data, respectively. In this section we reinterpret this search in the context of the model in \cref{sec:model} to set limits on the vector leptoquark mass as a function of $\beta_{23}^L$. In our limits we use the CMS data. Since ATLAS data presents a (small) upper fluctuation with respect to the SM background, a combination of ATLAS and CMS data yields slightly weaker limits than CMS data alone (see e.g.~\cite{Greljo:2018tzh}).

\begin{table}[t!]
\setlength{\tabcolsep}{20pt}
\renewcommand{\arraystretch}{1.5}
\centering
\begin{tabular}{cc}
\hline
Particle selection 1& No events with an electron ($p_T^e>20$~GeV, $|\eta_e|<2.4$) \\
Particle selection 2&  No events with a muon ($p_T^\mu>20$~GeV, $|\eta_\mu|<2.5$)\\
$\tau_h$ $p_T$ & $p_T^{\tau_{h}}>80$~GeV\\
Missing energy & $E_T^{\,\rm miss}>200$\\
$p_T$ vs missing energy & $0.7<p_T^{\tau_{h}}/E_T^{\,\rm miss}<1.3$\\
$\phi$ & $|\Delta\phi(p_T^{\tau_{h}},p_T^{\,\rm miss})|>2.4$~rad\\
\hline
\end{tabular}
\caption{Summary of the experimental cuts for the CMS $\tau_h\,\nu$ search~\cite{Sirunyan:2018lbg}.}
\label{tab:cutsDYtanuCMS}
\end{table}

\subsubsection{Search strategy}
\label{sec:pp2tanu_strategy}	

We compute the NP contribution to the $pp\to\tau_h\nu$ process, including 
the interference with the SM, using {\tt Madgraph5\_aMC@NLO v2.6.3.2}~\cite{Alwall:2014hca}, 
with the {\tt NNPDF23\_lo\_as\_0119\_qed} PDF set~\cite{Ball:2012cx}. 
Hadronization of the $\tau$ final state is done with {\tt Pythia 8.2}~\cite{Sjostrand:2014zea} 
using the CUETP8M1 set of tuned parameters~\cite{Khachatryan:2015pea}. The detector response is simulated 
using {\tt Delphes~3.4.1}~\cite{deFavereau:2013fsa}. The CMS Delphes card has been 
modified to satisfy the object reconstruction and identification requirements, 
in particular we include the $\tau$-tagging efficiencies quoted in the 
experimental search~\cite{Sirunyan:2018lbg}. 

After showering and detector simulation, we apply the selection cuts specified in \cref{tab:cutsDYtanuCMS} using {\tt MadAnalysis~5 v1.6.33}~\cite{Dumont:2014tja}. As a cross-check we have generated the SM Drell-Yan $pp\to\tau\nu$ background and compared our results to the one quoted by CMS. A good agreement is found between the two samples, within $20\%$ of the quoted central values.

 After passing through selection cuts, the resulting events are binned according to their total transverse mass,
\begin{align}
m_T^{\rm tot}\equiv\sqrt{2p_T^{\tau_{h}}p_T^{\rm\,miss}[1-\cos\Delta\phi(\vec p_T^{\;\tau_{h}},\vec p_T^{\rm\;miss})]}\,,
\end{align}
with $p_T^{\tau_h}$ and $p_T^{\rm miss}$ being, respectively, the transverse momenta of the visible decay products of the $\tau$ and the missing transverse momentum in the reconstructed event. We compare our binned events with the data and background estimates presented in fig.~3 (left) of~\cite{Sirunyan:2018lbg}. As we did in \cref{sec:pp2tata}, we consider all the available bins in the $m_T^{\rm tot}$ distribution, treating their errors as uncorrelated.  For the statistical analysis we use the modified frequentist $\textrm{CL}_s$ method~\cite{Read:2002hq} computed with the {\tt ROOT}~\cite{Brun:1997pa} package {\tt Tlimit}~\cite{Junk:1999kv}. In the determination of the limit, we include a systematic uncertainty of $20\%$ in the NP signal to account for possible uncertainties related to the PDF, tau hadronization, detector simulation and unaccounted NLO corrections.

\begin{figure}[t!]
  \begin{center}
    \begin{tabular}{cc}
      \includegraphics[height=0.45\textwidth]{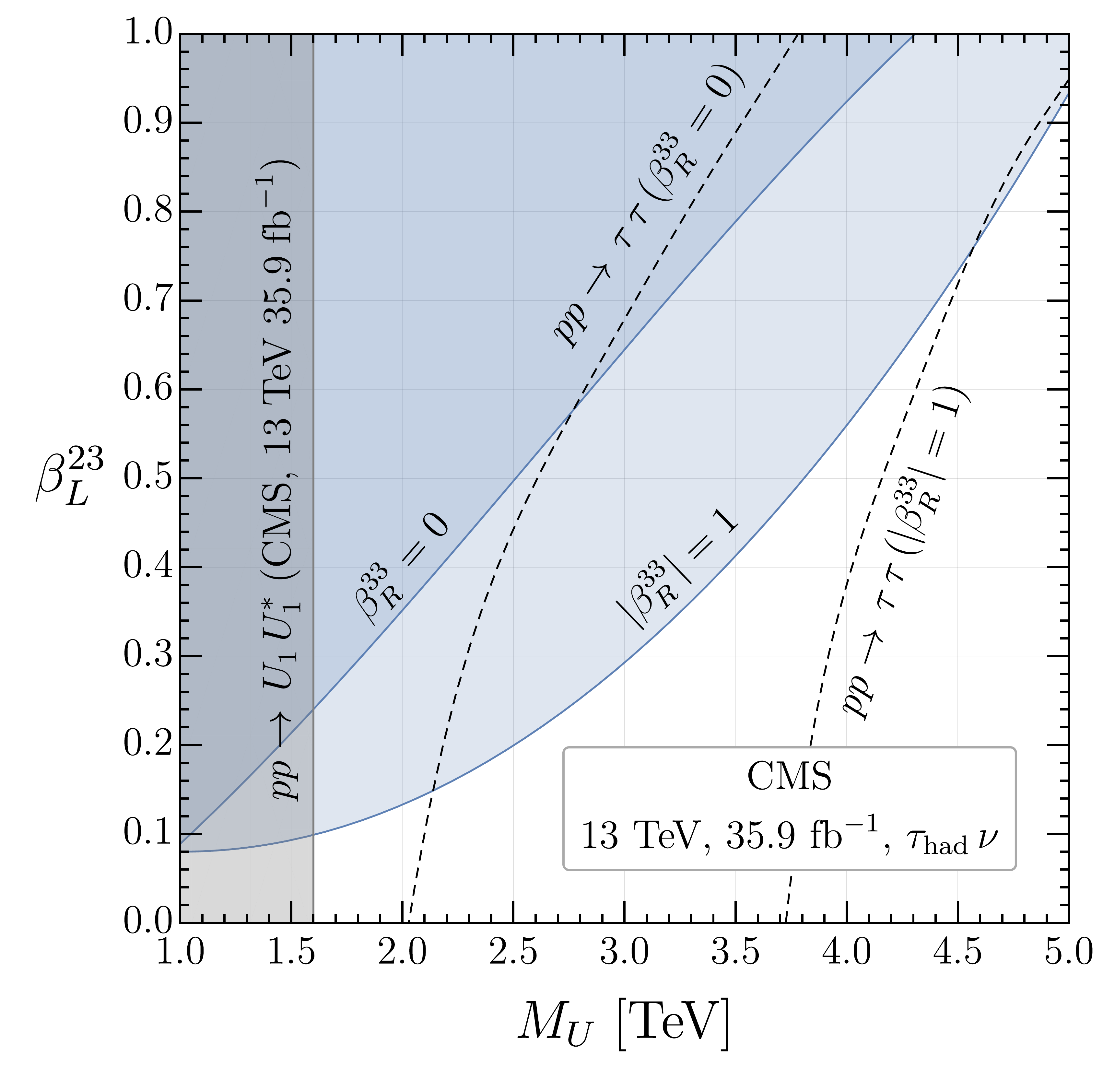}
    \end{tabular}
  \end{center}
\caption{Exclusion limits from the $pp\to\tau\nu$ search in the $(\beta_L^{23},M_U)$ plane for different values of the coupling $\beta_R^{33}$. We fix $\beta_L^{33}=1$, $g_U=3$ and the leptoquark width to its natural value. The corresponding limits from $pp\to\tau\tau$ and pair-production, using the same parameter points, are overlaid.}
\label{fig:LQ_tanu_excl}
\end{figure}	

\subsubsection{Limits on the \texorpdfstring{$U_1$ leptoquark}{leptoquark}}
\label{sec:pp2tanu_U1}	

For this search, we fix $\beta_L^{33}=1$ and consider two different benchmarks for the right-handed coupling, $|\beta_R^{33}|=0,\,1$.  In this case the relative sign between $\beta_R^{33}$ and $\beta_L^{33}$ is not observable in this channel.  Since the leptoquark width plays a marginal role, we fix it to its natural value. We furthermore set $g_U=3$.

We compute exclusion limits for the vector leptoquark in the $(\beta_L^{23},M_U)$ plane, see~\cref{fig:LQ_tanu_excl}. For comparison, we overlay the corresponding limits from $pp\to\tau\tau$ (see~\cref{fig:LQ_excl} (right)) and pair-production limits. As can be seen, these limits give complementary information to those presented in~\cref{sec:pp2tata_U1}, offering more stringent limits only when the $\beta_L^{23}$ coupling becomes large. Analogous limits for the case $\beta_R^{33}=0$ have already been derived in the past literature~\cite{Greljo:2018tzh}; we find good agreement between these limits and the ones quoted here. Interestingly, and as happens in the $pp\to\tau\tau$ search, the exclusion bounds get significantly affected by non-zero values of  $\beta_R^{33}$.  The different shapes in the exclusion bands can be understood from the fact that, for $|\beta_R^{33}|=1$, the dominant partonic process is $bc\to\tau\nu$, whose cross section scales as $\sigma_{bc\to\tau\nu}\sim|\beta_L^{23}|^2/M_U^4$ in the EFT limit. On the contrary, for $\beta_R^{33}=0$, the relative contribution from $sc$ production, for which $\sigma_{sc\to\tau\nu}\sim(|\beta_L^{23}|/M_U)^4$, is important and even becomes dominant for medium-size values of $\beta_L^{23}$.

\subsection{\texorpdfstring{$pp\to\tau\mu$}{Tau-muon} search}
\label{sec:pp2tamu}	

The ATLAS collaboration has published a search for heavy particles decaying into different-flavour dilepton pairs using $36.1~\textrm{fb}^{-1}$~\cite{Aaboud:2018jff} of 13~TeV data. In this section we recast the ATLAS data and reinterpret the collider bounds in terms of the model in \cref{sec:model} to set limits on $\beta_{32}^L$ and $\zeta^{23}_\ell$, as a function of the leptoquark and $Z^\prime$ masses, respectively.

\subsubsection{Search strategy}
\label{sec:pp2tamu_strategy}	

We use {\tt Madgraph5\_aMC@NLO v2.6.3.2}~\cite{Alwall:2014hca} with the {\tt NNPDF23\_lo\_as\_0119\_qed} PDF set~\cite{Ball:2012cx} to compute the NP contribution to the $pp\to\tau\mu$ process. The output is passed to {\tt Pythia 8.2}~\cite{Sjostrand:2014zea} for tau hadronization and the detector effects are simulated with {\tt Delphes~3.4.1}~\cite{deFavereau:2013fsa}. The ATLAS Delphes card  has been adjusted to satisfy the object reconstruction and identification criteria in the search. In particular we have modified the muon efficiency and momentum resolution to match the \textit{High-$\mathit{p_T}$} muon operating point, and adjusted the missing energy reconstruction to account for muon effects. We have further included the $\tau$-tagging efficiencies quoted in the experimental search~\cite{Aaboud:2018jff}. 

After showering and detector simulation, we apply the selection cuts specified in \cref{tab:cutsDYtamuATLAS} using {\tt MadAnalysis~5 v1.6.33}~\cite{Dumont:2014tja}.  The resulting events are binned according to their dilepton invariant mass. Following the approach described by ATLAS~\cite{Aaboud:2018jff}, the tau momentum is reconstructed from the magnitude of the missing energy and the momentum direction of the visible tau decay products.  This approach relies on the fact that the momentum of the visible tau decay products and the neutrino momentum are nearly collinear. 

In order to validate our procedure, we have simulated the $Z^\prime$ signal quoted in the experimental search~\cite{Aaboud:2018jff}, finding good agreement between our signal and the one by ATLAS.

We compared our results with the binned invariant mass distribution in~\cite{Aaboud:2018jff}. Since the error correlations are not provided, we treat the bin errors as uncorrelated. We use the modified frequentist $\textrm{CL}_s$ method~\cite{Read:2002hq} to obtain $95\%$ CL limits. These limits are computed using the 
 {\tt ROOT}~\cite{Brun:1997pa} package {\tt Tlimit}~\cite{Junk:1999kv}. In the determination of those limits, we include a systematic uncertainty of $20\%$ for the NP signal to account for possible uncertainties related to the PDF, tau hadronization, detector simulation and unaccounted NLO corrections.

\begin{table}[t!]
\setlength{\tabcolsep}{20pt}
\renewcommand{\arraystretch}{1.5}
\centering
\begin{tabular}{cc}
\hline
Particle selection & One single $\tau$ and $\mu$, no electrons\\
$p_T$ & $p_T^{\tau_h}>65$~GeV, $p_T^{\mu}>65$~GeV\\
$\eta$ & $|\eta_{\tau_h}|<2.5$ excluding $1.37<|\eta_{\tau_h}|<1.52$; $|\eta_\mu|<2.4$\\
$\phi$ & $|\Delta\phi(\tau_h,\mu)|>2.7$~rad\\
$\Delta R$ & $\Delta R(\tau_h,\mu)>0.4$\\
\hline
\end{tabular}
\caption{Summary of the experimental cuts for the ATLAS $\tau_h\,\mu$ search~\cite{Aaboud:2018jff}.}
\label{tab:cutsDYtamuATLAS}
\end{table}

\subsubsection{Limits on the \texorpdfstring{$U_1$ }{}leptoquark}
\label{sec:pp2tamu_U1}	

Following a similar strategy as for the other channels, we fix $g_U=3$ and $\beta_L^{33}=1$, and take two benchmark values for the right-handed coupling $|\beta_R^{33}|=0,1$ (different sign choices for this parameter do not have an impact on the high-$p_T$ signal). Varying the leptoquark width only yields a subleading effect so we keep it fixed to its natural value.

We decouple the $Z^\prime$ and compute the exclusion limits for the vector leptoquark mass as a function of $\beta_L^{32}$, see~\cref{fig:tamu_excl} (left). As in previously analysed channels, the exclusion limits vary significantly for different values of $\beta_R^{33}$. We additionally overlay the corresponding exclusion limit obtained from the $pp\to\tau\tau$ search and searches for pair-production. The limits from $pp\to\tau\mu$ become stronger than those obtained from $pp\to\tau\tau$ only for large values of the $\beta_L^{32}$ parameter, especially in the case when $|\beta_R^{33}|=1$. 

The limits presented here offer complementary constraints to the ones obtained from low-energy flavour observables. Indeed, one can establish a one-to-one correspondence between this search and the experimental limits from $\Upsilon(nS)\to\tau\mu$ decays. Using the expression in \cite{Kumar:2018kmr} we get
\begin{align}
\mathcal{B}(\Upsilon(2S)\to\tau\mu)\approx 2\cdot 10^{-6}\times\left(\frac{g_U}{3}\right)^4\times\left(\frac{1\,\mathrm{TeV}}{M_U}\right)^4\times|\beta_L^{32}|^2\,.
\end{align}
This is to be compared to the current experimental limit: $\mathcal{B}(\Upsilon(2S)\to\tau\mu)_{\rm exp}<3.3\cdot10^{-6}$ ($90\%$~CL). Interestingly, we find the current bounds from high-$p_T$ data (see~\cref{fig:tamu_excl} left) to be much more constraining than those from its low-energy counterpart.\footnote{Current limits from loop-mediated transitions, such as $\tau\to\mu\gamma$, offer stronger bounds in certain UV completions~\cite{Bordone:2018nbg}. However, these bounds are more sensitive to the details of the UV completion and are therefore less robust.} Future improvements on $pp\to\tau\mu$ searches can serve as a valuable probe of the leptoquark flavour structure.

\subsubsection{Limits on the \texorpdfstring{$Z^\prime$}{Z'}}
\label{sec:pp2tamu_Zp}	

\begin{figure}[t!]
  \begin{center}
    \begin{tabular}{cc}
      \includegraphics[height=0.45\textwidth]{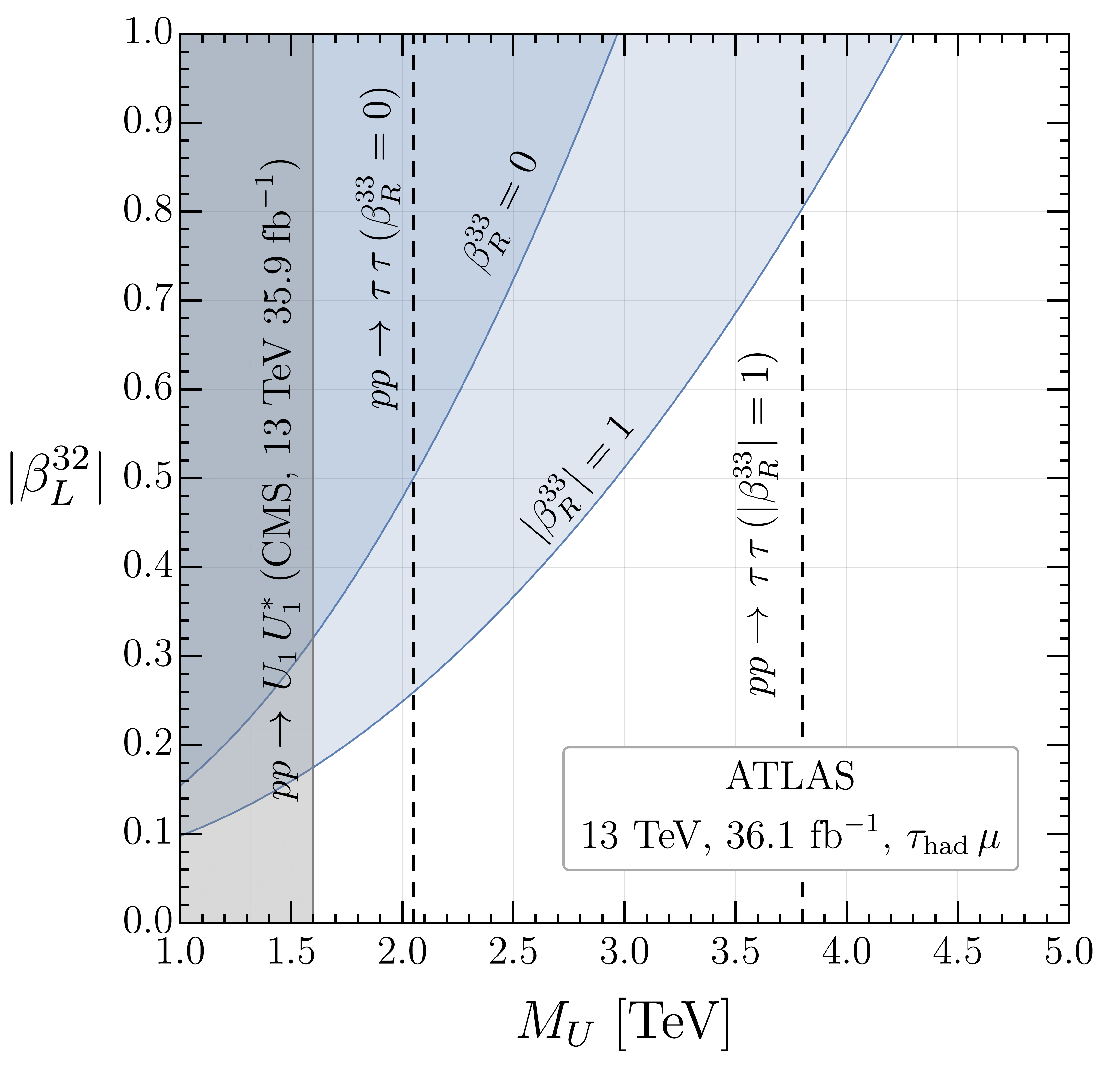} &       
      \includegraphics[height=0.45\textwidth]{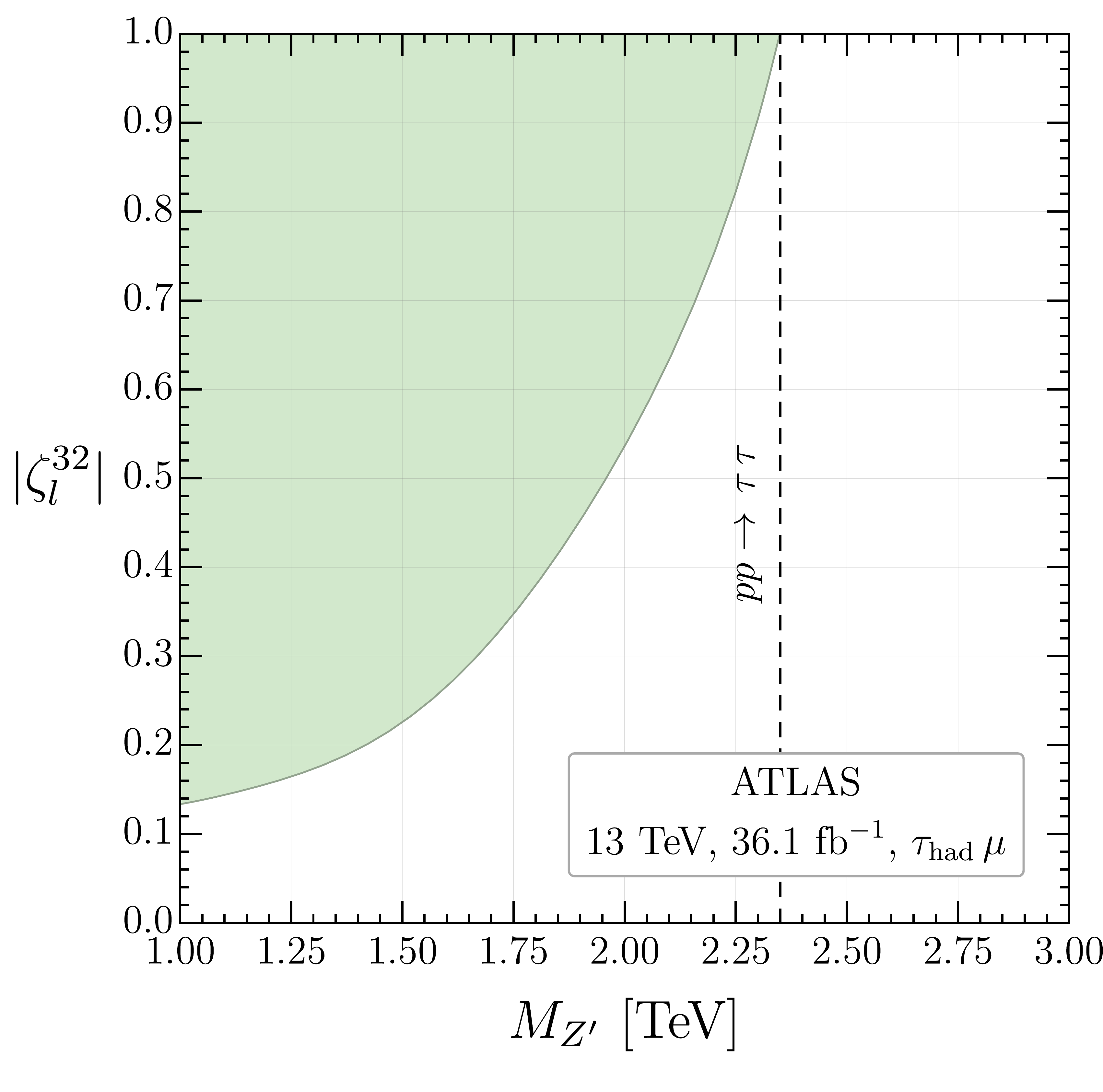}
    \end{tabular}
  \end{center}
\caption{$95\%$~CL exclusion limits from the $pp\to\tau\mu$ search. Left: $U_1$ limits in the $(\beta_L^{32},M_U)$ plane for different values of the coupling $\beta_R^{33}$. We fix $\beta_L^{33}=1$, $g_U=3$ and the leptoquark width to its natural value. Right: $Z^\prime$ limits in the $(\zeta_L^{32},M_{Z^\prime})$ plane, taking the natural width for the $Z^\prime$ and fixing $g_{Z^\prime}=3$. For comparison, the bounds from $U_1$ pair-production and from $pp\to\tau\tau$ are also shown.}
\label{fig:tamu_excl}
\end{figure}

We finally comment on the limits on the $Z^\prime$, decoupling the leptoquark. We fix $g_{Z^\prime}=3$, $\zeta_{q,u,d}^{33} = \zeta_{\ell,e}^{33} =1$ and set the $Z^\prime$ width to its natural value. Limits on the $Z^\prime$ mass as a function of $\zeta_\ell^{32}$ are shown in~\cref{fig:tamu_excl} (right). As in the leptoquark case, we overlay the corresponding limits on the $Z^\prime$ mass extracted from $pp\to\tau\tau$. As can be seen, these limits are always stronger than those from the present search, irrespective of the value of $\zeta_\ell^{32}$.

\subsection{\texorpdfstring{$pp\to t\bar t$}{Ditop} search}
\label{sec:pp2tt}	
We finally turn our attention to searches in the ditop final state, which is subject to NP effects from $s$ channel colorons and $Z'$ bosons. {We focus our analysis on the coloron since the bounds from this channel on the $Z^\prime$ 
are weaker than the ones reported in \cref{sec:pp2tata_Zp}. 

\subsubsection{Search strategy}
We perform a recast of the ATLAS study~\cite{Aaboud:2018eqg}, using 36~fb$^{-1}$ of collected data. Since the data was unfolded in this work, we can compute parton-level predictions and directly compare them to the unfolded distributions provided in the reference study.

We choose to derive the constraints from the normalised parton-level differential cross-sections as a function of the $t\bar t$-invariant mass, shown in fig.~14(b) of~\cite{Aaboud:2018eqg}.  As in the other searches, we do not include possible error correlations between the bins in the invariant-mass distribution since they are not provided. Our signal predictions are derived by integrating the leading-order SM QCD partonic cross-sections $q\bar q\to t\bar t$ and $gg\to t\bar t$ and the NP contributions from coloron and $Z'$ over the parton distribution functions, employing the \texttt{NNPDF30\_nlo\_as\_0119} PDF set~\cite{Ball:2012cx} and fixing the factorisation and renormalization scale to the center of the corresponding $t\bar t$-invariant mass bin. We use the running strong coupling constant as provided by the PDF set. The only cut applied is on transverse momentum of either top quark: $p_T^{t}>500~\mathrm{GeV}$. Note that our reference study places the cuts as $p_T^{t,1}>500~\mathrm{GeV}$ on the leading top, and $p_T^{t,2}>350~\mathrm{GeV}$ on the subleading one. For a fully exclusive, partonic $t\bar t$ final state, $p_T^{t,1}=p_T^{t,2}$ and hence the second cut does not influence our calculation. However, the unfolded distributions are derived from data which employ this slightly milder cut, leading to slight deviations in bins of lower invariant mass. We therefore drop the bins $m_{t\bar t}<1.2~\mathrm{TeV}$ and then find excellent agreement with the SM predictions presented in the ATLAS study. While the analysis also provides unfolded spectra differential in $p_T$ and various other kinematic observables, we find the invariant mass spectrum to be the most constraining distribution. We therefore focus solely on the invariant mass spectrum and do not consider searches in the angular spectra.

\subsubsection{Limits on the coloron}

We are now ready to present the constraints on the various parameters related to the coloron. Throughout this section
we set $\kappa_q^{33}=1$ and $\kappa_q^{ll}=\kappa_u^{ll}=\kappa_d^{ll}=-(g_s/g_{G'})^2$, and we fix $g_{G'}=3$, unless
otherwise stated.

\begin{figure}[t!]
\centering
\includegraphics[height=0.4\textwidth]{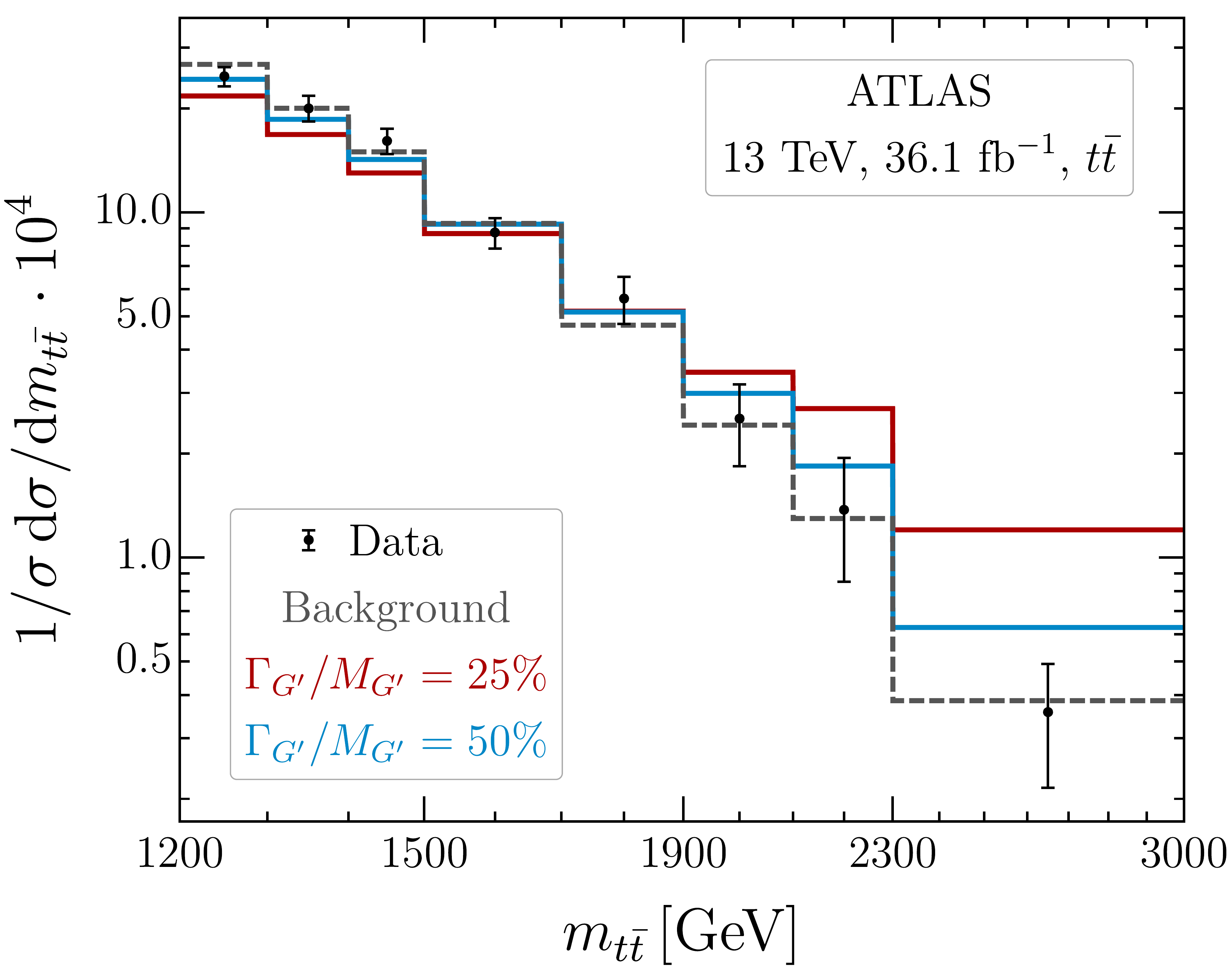} 
\caption{Illustration of the coloron signal in the ditop final state for $M_{G'}=2.5$~TeV, $\kappa_q^{ll}=0$,
and two reference values for the width, compared to the data from~\cite{Aaboud:2018eqg}. }
\label{fig:col_spec}
\end{figure}	
Before discussing each plot by itself, a few general comments are in order. Since the search is in a normalised spectrum, it is mostly sensitive to signals that create a shape sufficiently distinct to the background. A distinct shape in this case means a change in the spectrum that is unaffected by normalisation, meaning a peak or a change in the overall slope. 
Uniform shifts in the spectra (originating from resonances with both very high and very low masses, or large widths\footnote{A very wide resonance also leads to a suppression in the overall signal cross-section, further decreasing the constraining power of the search.}) are washed out by the overall normalisation.
As a result,  the strongest bounds are obtained when the coloron can be produced nearly on-shell and the width is moderate. 
For example, in \cref{fig:col_spec} we show the signal of two parameter points with $\kappa_q^{ll}=0$, $M_{G'}=2.5$~TeV and different choices of the coloron width.  We see that the narrower coloron results in a larger change in the slope.

In \cref{fig:col_excl}, we show exclusion regions for the coloron with its natural width and with a width enhanced by a factor of two. In the left panel, the exclusion limits in the $(g_{G'},M_{G'})$ plane are shown for the natural width and twice this value. An interesting feature of these exclusion regions is that the boundaries bend towards smaller masses for larger couplings. This can be understood by the fact that while the cross section grows with the coupling, so does the width. For the reasons discussed above, the search then loses sensitivity to the resulting signal.

In the right panel of \cref{fig:col_excl}, exclusions are shown for varying values of the coupling to left-handed light quarks $\kappa_{q}^{ll}$, keeping the right-handed couplings $\kappa_{u,d}^{ll}$ fixed. With larger couplings to the light quarks, the production of the coloron from valence quarks of the proton increases drastically. Since the valence quarks tend to carry more of the protons' momenta, they can produce the coloron closer to its mass shell, leading to a signal that the search 
can more easily discriminate from the background. If we were to set $\kappa_u^{ll}=\kappa_d^{ll}=\kappa_q^{ll}=0$, we would found only very mild bounds, in which case the pair-production search discussed in \cref{sec:pprod} outperforms this one. If $\kappa_q^{ll}$ is chosen to be positive, the bounds tend to be weaker due to interference between the NP and the SM contributions.
\begin{figure}[t!]
  \begin{center}
    \begin{tabular}{cc}
      \includegraphics[height=0.45\textwidth]{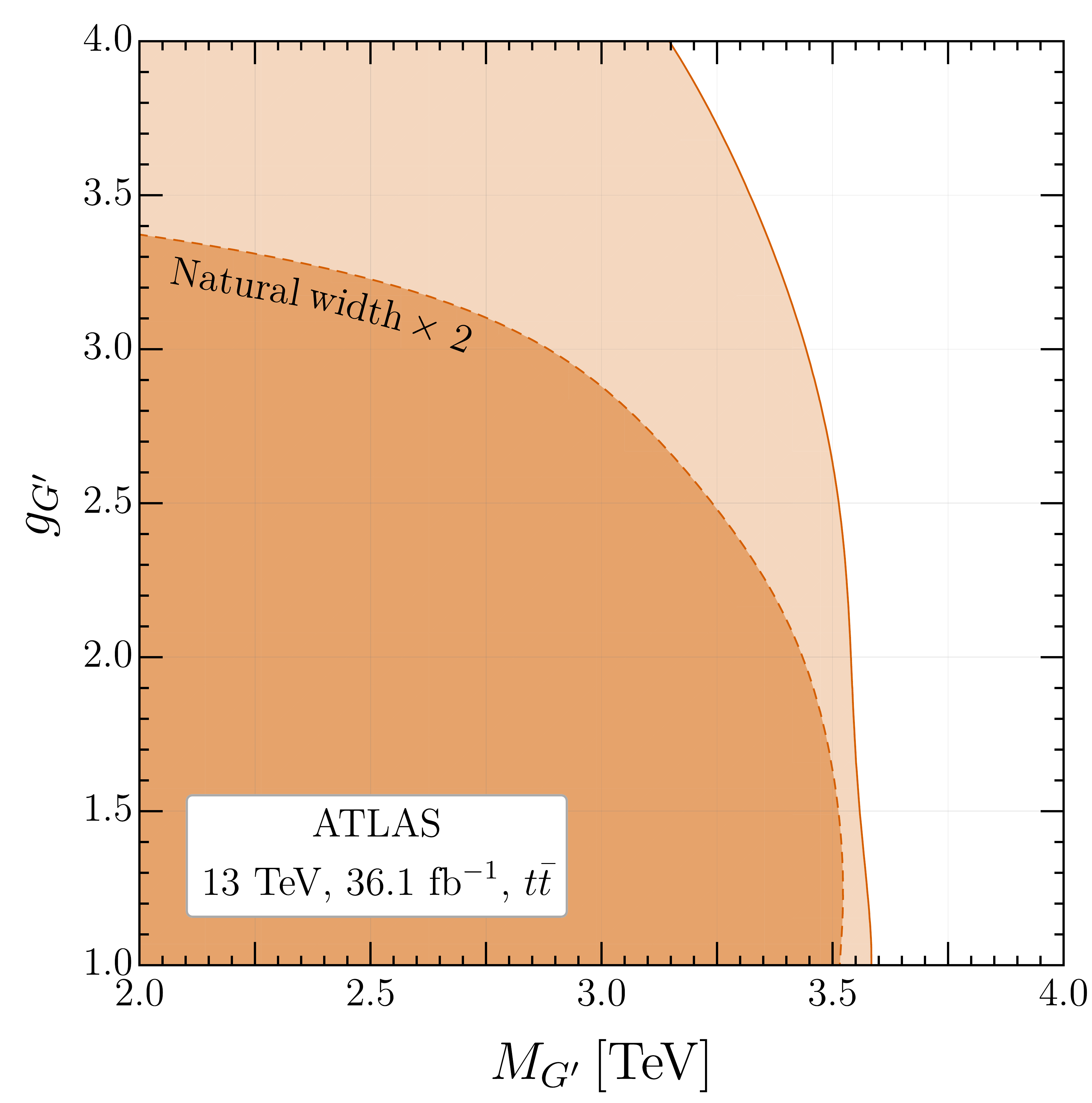} &   \includegraphics[height=0.45\textwidth]{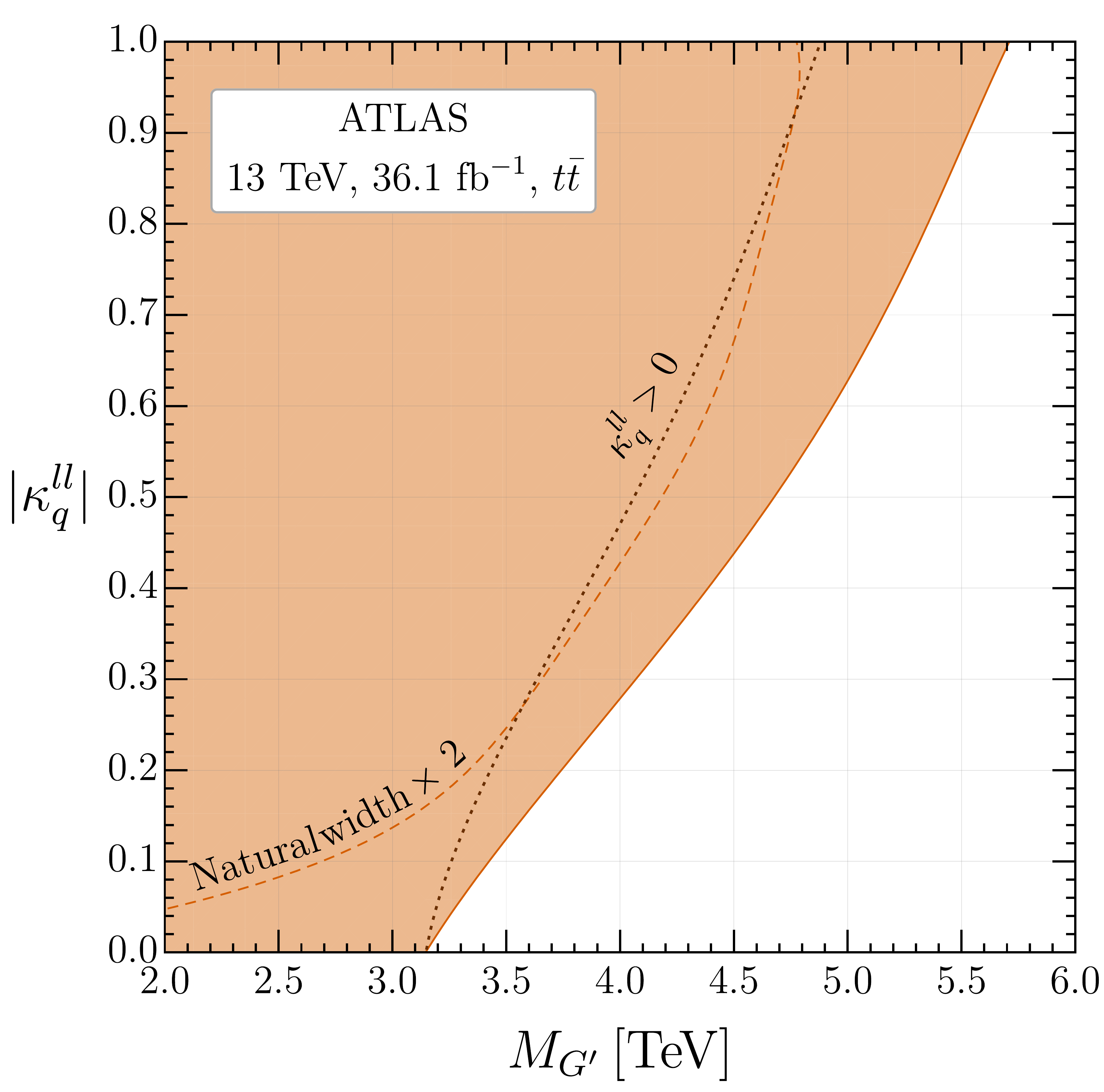}
    \end{tabular}
  \end{center}
\caption{Left: Exclusion plot for the $pp\to t\bar t$ search in the $(g_{G'},M_{G'})$ plane for the natural width (solid line) and twice the natural width (dashed line). Right: Exclusion limits on the coupling to light left-handed quarks $\kappa_q^{ll}$. The regions bounded by solid and dashed lines correspond to $\kappa_q^{ll}<0$ with the natural width and twice the natural width, respectively. The dashed line denotes the exclusion region for $\kappa_q^{ll}>0$.}
\label{fig:col_excl}
\end{figure}
\begin{figure}[t!]
  \begin{center}
    \begin{tabular}{cc}
      \includegraphics[height=0.45\textwidth]{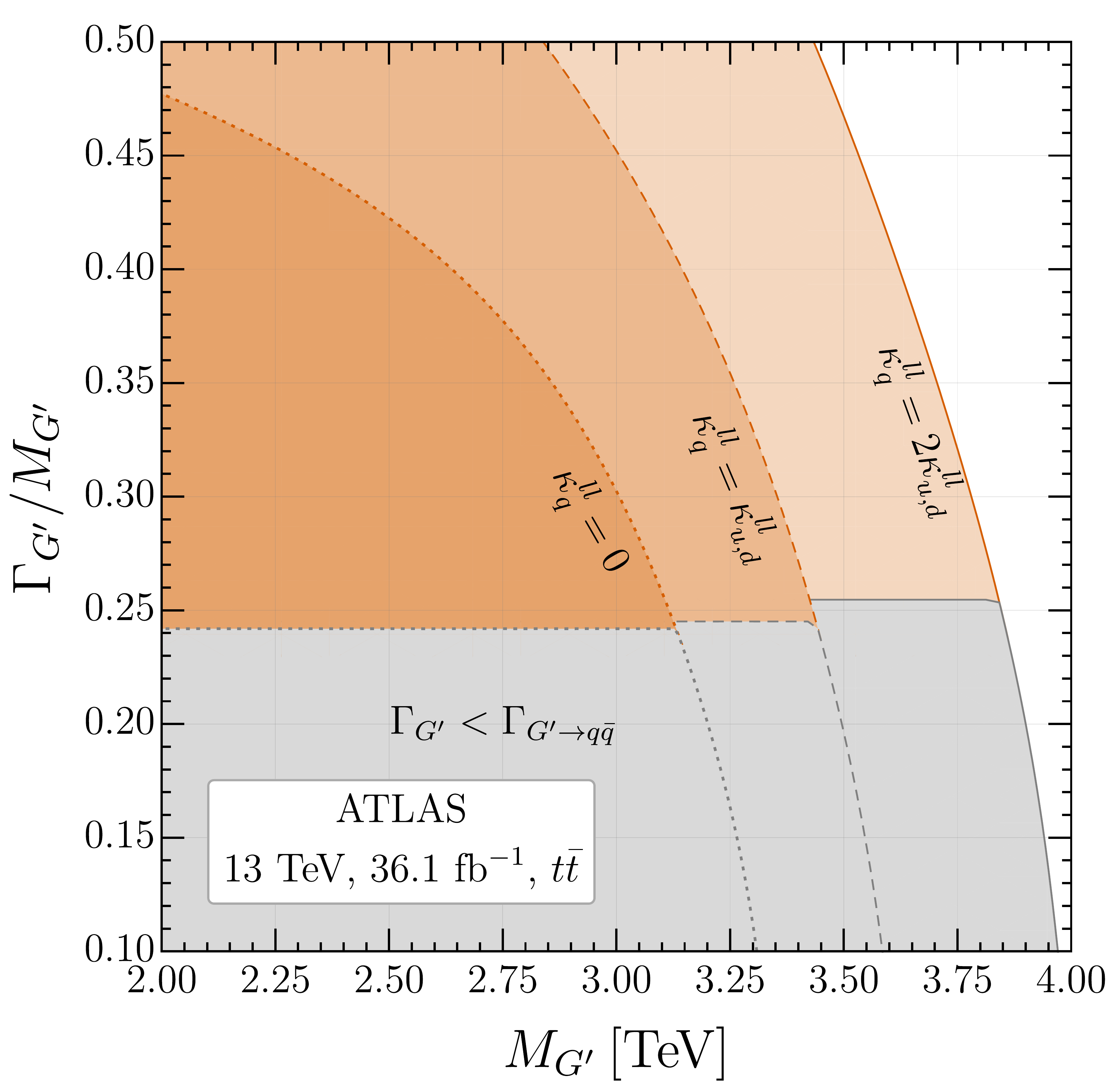} &   \includegraphics[height=0.45\textwidth]{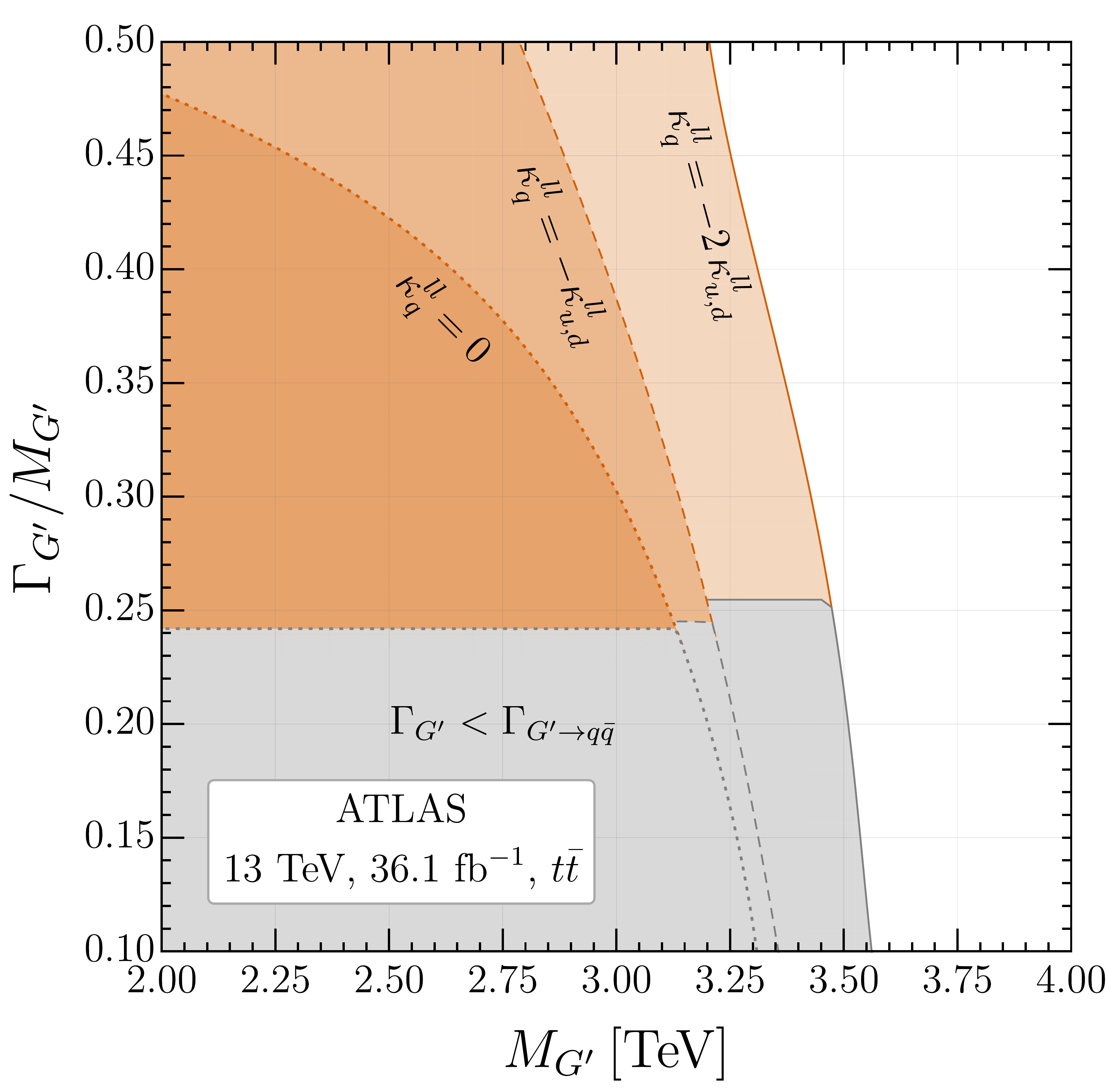}
    \end{tabular}
  \end{center}
\caption{Exclusion limits on the coloron for the $pp\to t\bar t$ search in the $(\Gamma_\mathrm{G'}/M_{G'},M_{G'})$ plane for different choices of the couplings to light left-handed quarks.}
\label{fig:col_excl2}
\end{figure}	
	

Finally, \cref{fig:col_excl2} shows exclusion limits with varying widths of the coloron. The different curves (solid, dashed, dotted) show various different choices of relations between the couplings to left- and right-handed light quarks. As expected, limits get weaker with increasing width of the resonance. When the sign of $\kappa_q^{ll}$ is chosen to be opposite of $\kappa_{u,d}^{ll}$, the bounds also become weaker for the same reason as discussed above. The grey bands denote the regions in which the floating width parameter is below the partial width to quarks. Note that for $\kappa_{G'}\neq 0$ the coloron can decay to two gluons, in which case the actual width would become significantly larger than the partial width to quarks alone.\footnote{In this case, the production cross section of the coloron would also be drastically increased, leading to much stronger bounds on its mass from this search.}

	\section{Conclusions}
	\label{sec:conclusions}	

The high-$p_T$ phenomenology of models predicting a TeV-scale $SU(2)_L$ singlet vector leptoquark which is 
able to account for the hints of  LFU violations observed in $B$-meson decays is quite rich. 
This is both because this exotic mediator can manifest itself in different final states accessible at the LHC, 
and also because this state cannot be the only TeV-scale exotic vector. As we have shown, the minimal 
consistent set of massive vectors comprising a $U_1$ also includes a coloron and a $Z^\prime$. 
In this paper we have presented a comprehensive analysis of the high-$p_T$ signatures  
of this set of exotic TeV states, deriving a series of bounds on their masses and couplings.

The results have been discussed in detail in the previous section and will not be repeated here. 
Here we limit ourself to summarise a few key messages, emphasising the novelties of our analysis 
compared to the results in the existing literature:
\begin{itemize}
\item In most of the relevant parameter space the most stringent bound on the leptoquark is obtained by the 
$pp\to \tau\tau$ process. In this channel a possible $O(1)$ right-handed coupling ($\beta_R^{33}$)
has a very large impact, as shown in \cref{fig:LQ_excl}. 
\item
A non-vanishing off-diagonal coupling of the $U_1$ to quarks  has a modest impact in $pp\to \tau\tau$,
provided  $|\beta_L^{23}| \lsim 0.2$ (as expected from a natural flavour structure), but a significantly 
larger impact in $pp\to \tau\nu$. However, the latter search remains subleading compared 
to $pp\to \tau\tau$ up to $|\beta_L^{23} | \lsim 0.8$  for 
for $|\beta_R^{33}|=1$ (or up to $|\beta_L^{23} | \lsim 0.6$ for $|\beta_R^{33}|=0$).
\item
For large non-vanishing off-diagonal coupling to leptons, a potentially interesting channel is 
$pp\to \tau\mu$. In the pure left-handed case, the bound from this channel is stronger than 
the one from $pp\to \tau\tau$ if $|\beta_L^{32}| \geq 0.5$ (see \cref{fig:tamu_excl}).
\item Taking $g_U=g_{Z^\prime}$ and assuming dominant third-generation coupling to fermions and small 
couplings to the light families, the constraints on the $Z^\prime$  mass are significantly weaker 
than those on the $U_1$ (see \cref{fig:MZpLQ_excl}). The combination of $U_1$ and $Z^\prime$ signals 
in $pp\to \tau\tau$ leads to a modest increase on the corresponding bounds, confined to a relatively narrow region of the
parameter space. 
\item
The bound on the coloron from $pp\to t\bar t$ is quite sensitive to the width of this state, and to the possible coupling 
to light quarks. Due to the increase of the width, the bounds become weaker at large couplings (see \cref{fig:col_excl2} (left)).
\end{itemize}

\begin{figure}[t!]
\centering
\includegraphics[width=0.5\textwidth]{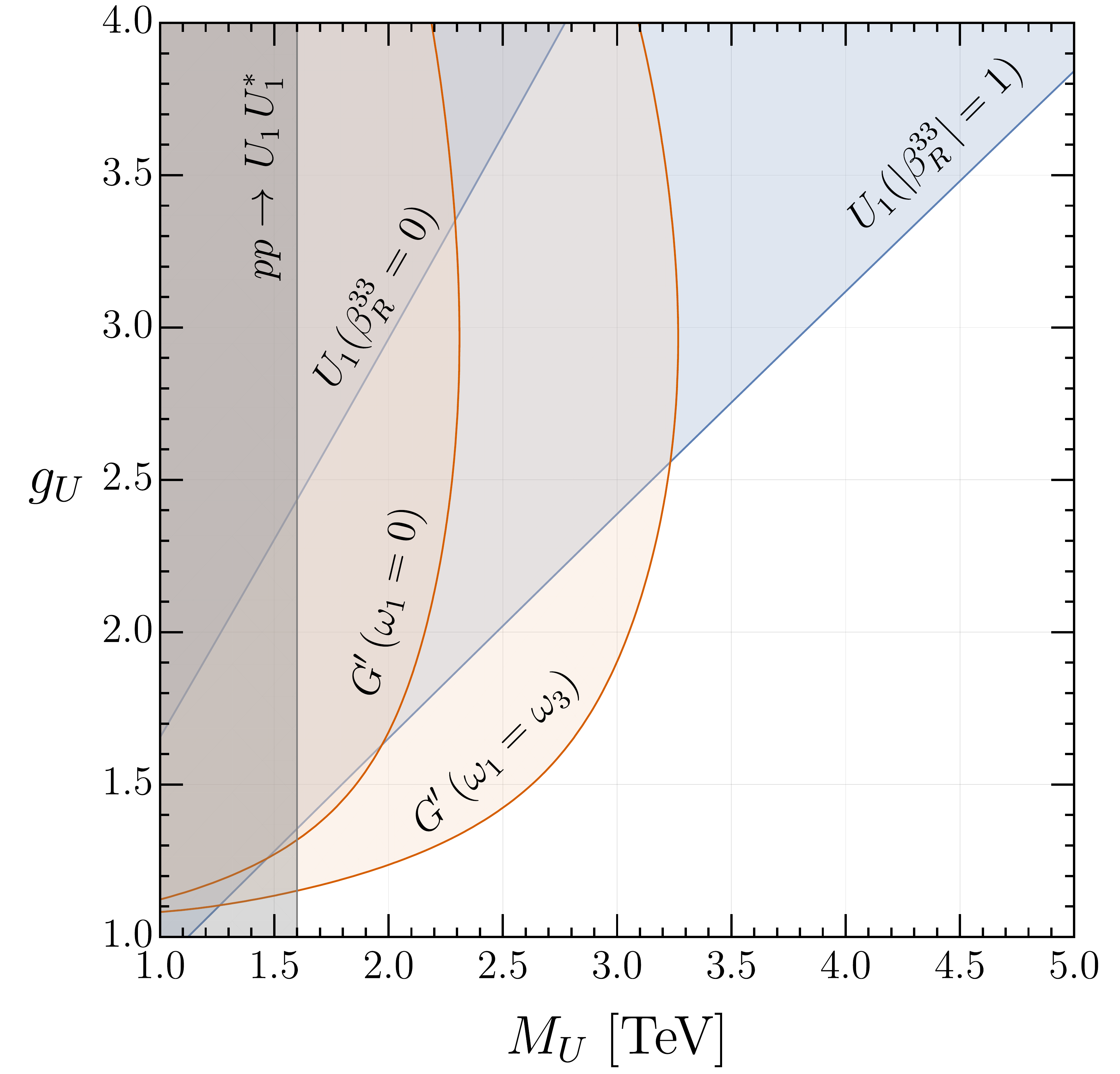} 
\caption{Leading $95\%$ CL exclusion limits for the $U_1$ and the coloron, shown in the $(g_U, M_U)$ plane assuming 
the relation between their masses and couplings following from the gauge symmetry  in \eqref{eq:Gntomin}  
and the breaking pattern assumed in~\cite{DiLuzio:2017vat,Bordone:2017bld}. See text for more details.}
\label{fig:LQ_Gp_excl}
\end{figure}	

The bounds we have obtained are very general and can be applied to a large class of models. One of the advantages 
of having analysed the three states together is the possibility of performing a direct comparison 
of the bounds obtained from the different mediators (via different processes) on the same model parameter space.
As an illustration of this fact, in \cref{fig:LQ_Gp_excl} we show a comparison of 
$U_1$ and coloron bounds in the $(g_U, M_U)$ plane, 
assuming the following relation between their masses and couplings
\begin{align}\label{eq:GpUrel}
M_{G^\prime}&=M_U\,\frac{g_U}{\sqrt{g_U^2-g_s^2}}\,\sqrt{\frac{2\,\omega_3^2}{\omega_1^2+\omega_3^2}}\,, & g_{G^\prime}&=\sqrt{g_U^2-g_s^2}\,.
\end{align}
This relation follows from the gauge symmetry  in \cref{eq:Gntomin}  assuming two breaking terms 
transforming as $\omega_3\sim(4,\bar 3)$ and $\omega_1\sim (4,1)$ under 
$SU(4)\times SU(3)^\prime$~\cite{DiLuzio:2017vat,Bordone:2017bld}. 
As can be seen, there is an interesting interplay between the two types of bounds, which changes according to the 
(model-dependent) ratio $\omega_1/\omega_3$. Once more, it is worth stressing the importance of the possible 
right-handed coupling of the $U_1$ (neglected in previous analyses): while the coloron  
sets the most stringent bounds on most of the parameter space for $\beta_R^{33}=0$, this is no-longer true for 
$\beta_R^{33}=O(1)$. This fact has relevant phenomenological consequences. For instance, the 
benchmark point corresponding to $20\%$ correction in $R(D)$ discussed after Eq.(\ref{eq:RD}) is excluded 
if $\beta_R^{33}=0$, while it is not yet excluded for $\beta_R^{33}=-1$.
More generally, given the minor role of the coloron bounds when $\beta_R^{33}=O(1)$, 
we can state that there is a more direct connection between high-$p_T$ physics and $B$-physics anomalies
in models with a large right-handed leptoquark coupling.

\section*{Acknowldgements}
\label{sec:acknowledgements}	
We thank Claudia Cornella and Admir Greljo for useful comments on the manuscript. JFM is grateful to the Mainz Institute for Theoretical Physics (MITP) for its hospitality and its partial support while this work was being finalised. We would like to thank Hubert Spiesberger for allowing us to use the THEP Cluster in Mainz for parts of this study. This research was supported in part by the Swiss National Science Foundation (SNF) under contract 200021-159720.

\vspace{0.5cm}

\bibliographystyle{JHEP}

{\footnotesize
\bibliography{paper}
}

\end{document}